\documentclass[showpacs,aps,prd,reprint,superscriptaddress,nofootinbib,longbibliography,preprintnumbers]{revtex4-1}
\usepackage[colorlinks=true, pdfstartview=FitV, linkcolor=magenta,citecolor=blue, urlcolor=magenta,
bookmarks=true, bookmarksnumbered=true, breaklinks]{hyperref}
\usepackage[dvipdfmx]{graphicx}

\usepackage{here}
\usepackage{url}
\usepackage{amsmath,amssymb,bm,color,longtable,mathrsfs,amsfonts,slashed}

\newcommand{\Slash}[1]{{\ooalign{\hfil/\hfil\crcr$#1$}}}

\begin{document}
\begin{flushright}
\end{flushright}

\title{Enhancement of axial anomaly effects in hot two-color QCD: \\
FRG approach in the linear sigma model}

\author{Gergely Fej\H{o}s}
\affiliation{Institute of Physics and Astronomy, E\"otv\"os University, 1117 Budapest, Hungary,}
\affiliation{RIKEN Center for Interdisciplinary Theoretical and Mathematical Sciences (iTHEMS), Wako, Saitama 351-0198, Japan}
\author{Daiki Suenaga}
\affiliation{Kobayashi-Maskawa Institute for the Origin of Particles and the Universe, Nagoya University, Nagoya 464-8602, Japan}
\affiliation{Research Center for Nuclear Physics, Osaka University, Ibaraki 567-0048, Japan}

\date{\today}

\begin{abstract}
We investigate the thermal properties of hadrons in two-color quantum chromodynamics (QC$_2$D) using the functional renormalization group method, with particular focus on modifications of the $U(1)$ axial anomaly effects. The hadrons are described by a linear sigma model based on the Pauli-G\"{u}rsey $SU(4)$ symmetry, which incorporates both low-lying $0^\pm$ mesons and diquark baryons. We find that all quartic couplings are comparably suppressed when physical values of the pion mass and decay constant are taken as inputs, for which a reasonably smooth chiral symmetry restoration at finite temperature is reproduced. Consequently, mass differences among chiral partners remain small. Despite these tiny mass differences, mass degeneracies of chiral partners in the hot medium are clearly demonstrated, consistent with chiral symmetry restoration. Moreover, we find that the couplings responsible for the $U(1)$ axial anomaly are enhanced upon entering the finite-temperature regime. Baryonic fluctuations also provide sizable contributions to these enhancements. Finally, the fate of the topological susceptibility in the hot QC$_2$D medium is examined.
 \end{abstract}


\maketitle

\section{Introduction}
\label{sec:Introduction}

The $U(1)$ axial anomaly, being responsible for the origin of a large $\eta'$ meson mass, is one of the most important symmetry properties of quantum chromodynamics (QCD). Unlike the breakdown of chiral symmetry, the $U(1)_A$ anomaly does not arise from spontaneous symmetry breaking but is caused by topological fluctuations of gluons, i.e., instantons~\cite{tHooft:1976rip}. Delineating the effects of the $U(1)_A$ anomaly in the hadronic regime helps illuminate interactions that cannot be fully explained by chiral symmetry alone.

In the hadronic regime, within chiral effective models, such as the linear sigma model (LSM), the $U(1)_A$ anomaly effects can be well incorporated by the so-called Kobayashi-Maskawa--'t Hooft (KMT) determinant~\cite{Kobayashi:1970ji,Kobayashi:1971qz,tHooft:1976snw,tHooft:1976rip}. The coefficient of the KMT term is proportional to the instanton density in the underlying theory, according to a weak coupling calculation on an instanton background. Thus, one can easily predict the disappearance of the anomaly, and, therefore the determinant contribution of KMT in the effective model, at the high temperature and/or density limit, due to the Debye screening of electric gluons~\cite{Gross:1980br,Rapp:1999qa}. At intermediate temperature ranges, on the other hand, the $U(1)_A$ anomaly effects on hadrons are not well understood~\cite{Fejos:2015xca}.

 In Refs.~\cite{Fejos:2016hbp,Fejos:2021yod}, based on a functional renormalization group (FRG) treatment for the $N_f=2+1$ LSM, a scenario where the $U(1)_A$ anomaly effects are strengthened at finite temperature due to mesonic fluctuations was proposed. Motivated by this finding, in this paper, we investigate the fate of the anomaly effects at finite temperature in two-color QCD (QC$_2$D), in order to theoretically confirm whether the anomaly enhancement is universal to the number of colors. In what follows, we apply the FRG method to a LSM of QC$_2$D with two flavors ($N_f=2$), which was introduced in Ref.~\cite{Suenaga:2022uqn}. This version of the LSM has been developed mainly to investigate the cold and dense properties of QC$_2$D matter extensively~\cite{Kawaguchi:2023olk,Kawaguchi:2024iaw,Suenaga:2023xwa} (for a review, see~\cite{Suenaga:2025sln}).

The FRG, as a mathematical tool, is one of the most promising functional methods in quantum field theory, capable of capturing essential features of strongly interacting quantum systems \cite{Berges:2000ew,Dupuis:2020fhh}. To the best of our knowledge, there have been no attempts using the FRG method to describe the thermodynamics of QC$_2$D with both mesonic and baryonic degrees of freedom included. In addition to exploring the thermal behavior of the $U(1)_A$ anomaly, this paper also aims to establish an FRG-based framework capable of accounting for the thermodynamic properties of two-color QCD in low-energy regimes~\cite{Fejos:2025nvd}.

In QC$_2$D, (anti)diquarks composed of two (anti)quarks are counted as color singlet hadrons. These (anti)diquark baryons can be treated on an equal footing with mesons due to the pseudoreality of the color group $SU(2)$: ${\bm 2}\simeq \bar{\bm 2}$. In terms of chiral representation, this property is reflected by an extension of chiral symmetry; in QC$_2$D, $SU(N_f)_L\times SU(N_f)_R$ chiral symmetry is extended to an $SU(2N_f)$ one, where (anti)diquark baryons and mesons belong to the same multiplet~\cite{Kogut:1999iv,Kogut:2000ek}. This extended symmetry is often referred to as the Pauli-G\"{u}rsey $SU(2N_f)$ symmetry~\cite{Pauli:1957voo,Gursey:1958fzy}.

Since the LSM in QC$_2$D is constructed upon the Pauli-G\"{u}rsey symmetry, one can naturally incorporate baryonic contributions in addition to mesonic ones in evaluating the $U(1)_A$ anomaly effects. The diquark contributions cannot be easily captured in ordinary three-color QCD, and, thus we will pay particular attention to separating baryonic and mesonic contributions in order to gain insights into fluctuations specific to QC$_2$D.

In-medium properties of QC$_2$D are currently being investigated by first-principles lattice simulations, especially due to the absence of the sign problem at finite baryon chemical potential~\cite{Boz:2019enj,Buividovich:2020dks,Astrakhantsev:2020tdl,Iida:2024irv,Braguta:2023yhd}. Therefore, in the future, one may expect {\it numerical experiments} on the lattice, which are capable of testing our current findings. In QC$_2$D, there is no {\it a priori} physical point. In the present analysis, we take physical values of the pion mass and the pion decay constant, while $\eta$ mass is considered as a free parameter, in order to investigate the $U(1)_A$ anomaly effects in a wider range. We also examine the fate of the anomaly effects as well as hadron mass modifications in a hot medium to provide useful information for future numerical experiments. 

This article is organized as follows. In Sec.~\ref{sec:Model}, we briefly review the emergence of the Pauli-G\"{u}rsey $SU(4)$ symmetry in QC$_2$D with $N_f=2$ and introduce our LSM. In Sec.~\ref{sec:FlowEquation}, our approach, based on the FRG framework, is explained, and RG flow equations for the couplings are derived. Then, in Sec.~\ref{sec:Result}, we present numerical results on the anomaly coefficients, hadron masses, and the topological susceptibility at finite temperature, by solving the flow equations. Section~\ref{sec:Baryons} is devoted to detailed analyses of baryonic contributions. Finally, Sec.~\ref{sec:Conclusions} is devoted to the conclusions.

\section{Two-flavor LSM in QC$_2$D}
\label{sec:Model}

In this section, we briefly review the two-flavor LSM in QC$_2$D introduced in Ref.~\cite{Suenaga:2022uqn}.

In QC$_2$D with $N_f=2$, the Lagrangian of quarks interacting with gluons is concisely expressed by a quartet of $\Psi =(\psi_R,\tilde{\psi}_L)^T= (u_R,d_R,\tilde{u}_L,\tilde{d}_L)^T$ with $\tilde{\psi}\equiv\sigma^2\tau_c^2\psi^* $ ($\sigma^a$ and $\tau_c^a$ are the Pauli matrices inhabiting spinor and color spaces, respectively) as
\begin{eqnarray}
{\cal L}_{\rm QC_2D} &=& \bar{\psi}i\Slash{D}\psi \nonumber\\
&=& \Psi^\dagger i\sigma^\mu D_\mu\Psi\ ,\label{LQC2D}
\end{eqnarray}
where in the second line we made use of the pseudorealities $\tau_c^a=\tau_c^2(\tau_c^a)^T\tau_c^2$ and $\sigma^i = -\sigma^2(\sigma^i)^T\sigma^2$~\cite{Kogut:1999iv,Kogut:2000ek}. In Eq.~(\ref{LQC2D}), the covariant derivative reads $ D_\mu={\partial}_\mu-ig_s{A}_\mu^a\tau_c^a$ with the strong coupling constant $g_s$ and $\sigma^\mu=({\bm 1},\sigma^i)$. Intuitively speaking, the second line in Eq.~(\ref{LQC2D}) implies that gluons cannot distinguish quarks and antiquarks due to the pseudoreality of the $SU(2)_c$ group and they can be described by a single multiplet $\Psi$ collectively. As a result, the Lagrangian with massless quarks possesses the so-called Pauli-G\"{u}rsey $SU(4)$ symmetry, as an extension of $SU(2)_L\times SU(2)_R$ chiral symmetry~\cite{Pauli:1957voo,Gursey:1958fzy}. 

Chiral effective models in QC$_2$D must respect the Pauli-G\"{u}rsey $SU(4)$ symmetry in order to accurately describe low-energy hadronic excitations. In particular, the LSM is based on the linear representation of this $SU(4)$ symmetry, in which hadron fields are arranged into a $4\times4$ matrix $\Sigma$. The transformation of $\Sigma$ under the $SU(4)$ group is given by an underlying quark bilinear: $\Sigma_{ij} \sim \Psi^T_j\sigma^2\tau_c^2\Psi_i$ ($i,j=1$ -- $4$). Hence, $\Sigma$ transforms as 
\begin{eqnarray}
\Sigma \to g \Sigma g^T\ ,\ \  {\rm with}\  \ g\in SU(4)\ .
\end{eqnarray}
In this representation the hadrons are assigned to be
\begin{eqnarray}
\Sigma = ({\cal S}^a-i{\cal P}^a) X^aE\ , \label{SigmaDef}
\end{eqnarray}
where ${\cal S}^a$ and ${\cal P}^a$ ($a=0$ -- $5$) are defined through the meson fields
\begin{eqnarray}
&& \eta = {\cal P}^0\ , \ \  \pi^\pm=\frac{{\cal P}^1\mp i{\cal P}^2}{\sqrt{2}} \ , \ \  \pi^0 = {\cal P}^3\ , \nonumber\\
&&\sigma = {\cal S}^0\ , \ \ a_0^\pm=\frac{{\cal S}^1\mp i{\cal S}^2}{\sqrt{2}} \ , \ \ a_0^0 = {\cal S}^3\ , \label{Mesons}
\end{eqnarray}
and baryon fields
\begin{eqnarray}
&& B = \frac{{\cal P}^5-i{\cal P}^4}{\sqrt{2}}\ , \ \ \bar{B} =  \frac{{\cal P}^5+i{\cal P}^4}{\sqrt{2}} \ , \nonumber\\
&& B' = \frac{{\cal S}^5-i{\cal S}^4}{\sqrt{2}}\ , \ \ \bar{B}' =  \frac{{\cal S}^5+i{\cal S}^4}{\sqrt{2}} \ . \label{Baryons}
\end{eqnarray}
Here, $B$ ($\bar{B}$) and $B'$ ($\bar{B}'$) are $0^+$ and $0^-$ (anti)diquark baryons, respectively. We note that mesons ($\bar{q}q$) and baryons ($qq$) are treated in a unified way reflecting the pseudoreality of $SU(2)$ color group: ${\bm 2}\simeq \bar{\bm 2}$. In Eq.~(\ref{SigmaDef}), $4\times 4$ matrices $X^a$ are generators belonging to the algebra of $G/H=SU(4)/Sp(4)$:
\begin{eqnarray}
X^{a=1,2,3} &=& \frac{1}{2\sqrt{2}}\left(
\begin{array}{cc}
\tau_f^a & 0 \\
0 &(\tau_f^a)^T \\
\end{array}
\right)\  , \nonumber\\
X^{a=4,5} &=& \frac{1}{2\sqrt{2}}\left(
\begin{array}{cc}
0& D_f^a \\
(D_f^a)^\dagger & 0 \\
\end{array}
\right) \ , \label{GeneratorX}
\end{eqnarray}
where $\tau_f^a$ ($a=1$ -- $3$) are the Pauli matrices acting on flavor space, $D^{a=4}=\tau_f^2$, and $D_f^{a=5}=i\tau_f^2$. Furthermore, $X^0={\bm 1}_{4\times4}/(2\sqrt{2})$, and 
\begin{eqnarray}
E = \left(
\begin{array}{cc}
 {\bm 0} & {\bm 1}\\
 -{\bm 1} & {\bm 0} \\
 \end{array}
 \right)
\end{eqnarray}
is the $4\times4$ symplectic matrix. 

It should be noted that, in QC$_2$D with $N_f=2$, the breaking pattern of chiral symmetry is $SU(4) \to Sp(4)$, leading to appearance of five Nambu-Goldstone bosons: three pions, a diquark baryon, and an antidiquark baryon.

With Eq.~(\ref{SigmaDef}), the Lagrangian of LSM takes the form of~\cite{Suenaga:2022uqn,Suenaga:2025sln}
\begin{eqnarray}
{\cal L}_{\rm LSM} = {\rm tr}[\partial_\mu\Sigma^\dagger\partial^\mu\Sigma] + {\cal L}_{\rm ex} -V\ , \label{LSMUV}
\end{eqnarray}
where the first term is a usual kinetic term. The second term
\begin{eqnarray}
{\cal L}_{\rm ex}  = \bar{c}{\rm tr}[\zeta^\dagger\Sigma + \Sigma^\dagger\zeta]
\end{eqnarray}
represents the explicit breaking of the Pauli-G\"{u}rsey $SU(4)$ symmetry, which generates finite masses of pion and $0^+$ diquark. Practically, the spurion field $\zeta$ is replaced by a quark mass $\zeta \to \langle\zeta\rangle=m_qE$, resulting in
\begin{eqnarray}
{\cal L}_{\rm ex}  \to \bar{c}m_q{\rm tr}[E^T\Sigma+\Sigma^\dagger E]\ . \label{Lex}
\end{eqnarray}
The potential part $V$ contains all possible terms up to ${\cal O}(\Sigma^4)$ preserving $SU(4)$ symmetry, but breaking $U(1)_A$ due to the axial anomaly. Defining invariants 
\begin{eqnarray}
I_1 &=& {\rm tr}[\Sigma^\dagger\Sigma] \ ,\nonumber\\
I_2 &=& {\rm tr}\big[(\Sigma^\dagger\Sigma-\frac{1}{4}{\rm tr}[\Sigma^\dagger\Sigma]{\bm 1})^2\big]\ , \nonumber\\
I_A &=& \frac{1}{2}{\rm tr}[\tilde{\Sigma}\Sigma + \tilde{\Sigma}^\dagger\Sigma^\dagger]  \ ,
\end{eqnarray}
with $\tilde{\Sigma}_{ij} =\frac{1}{2}\epsilon_{ijkl}\Sigma_{kl}$, $V$ reads
\begin{eqnarray}
\label{Vcl}
\!\!V = m^2I_1 + \lambda_1I_1^2+ \lambda_2I_2  + a I_A + c_1 I_A^2 + c_2I_1 I_A\ , \label{VUV}
\end{eqnarray}
where $I_A$ is responsible for the $U(1)_A$ anomaly, since $\Sigma \overset{U(1)_A}{\to} {\rm e}^{-i\theta_A}\Sigma {\rm e}^{-i\theta_A} = {\rm e}^{-2i\theta_A}\Sigma$. As for other anomalous terms, we may think of incorporating the KMT determinant $\det\Sigma + \det\Sigma^\dagger$ additionally. This, however, is not an independent invariant due to an identity of $\det \Sigma +\det \Sigma^{\dagger}=\frac14 I_A^2+\frac12 I_2-\frac18 I_1^2$: therefore, it must be omitted from the potential.\footnote{In Eq.~(\ref{VUV}) the terms containing $\lambda_1$ and $\lambda_2$ are rearranged as
\begin{eqnarray}
\lambda_1I_1^2+\lambda_2I_2 = \left(\lambda_1-\frac{\lambda_2}{4}\right)\big({\rm tr}[\Sigma^\dagger\Sigma]\big)^2 + \lambda_2{\rm tr}[(\Sigma^\dagger\Sigma)^2]\ . \label{Lambda1Lambda2} 
\end{eqnarray}
Thus, matching with the standard notation
\begin{eqnarray}
\bar{\lambda}_1\big({\rm tr}[\Sigma^\dagger\Sigma]\big)^2 + \bar{\lambda}_2{\rm tr}[(\Sigma^\dagger\Sigma)^2] \nonumber
\end{eqnarray}
yields $\bar{\lambda}_1=\lambda_1-\lambda_2/4$ and $\bar{\lambda}_2=\lambda_2$.}

We note that our LSM contains $12$ hadrons in total, listed in Eqs.~(\ref{Mesons}) and~(\ref{Baryons}), for which the Pauli-G\"{u}rsey $SU(4)$ symmetry of the potential~(\ref{VUV}) in the presence of the $U(1)_A$ anomaly is properly respected. This structure is mandatory to derive a closed set of flow equations as will be seen in Sec.~\ref{sec:FlowEquation}. An FRG analysis of QC$_2$D in the absence of the anomaly, including only the $\sigma$, $\pi$, $B$, and $\bar{B}$ as hadrons within the (Polyakov-)quark-meson-diquark model, was done in~\cite{Strodthoff:2013cua}.

In the following analysis, we will consider the quantum version of the classical theory of Eq.~(\ref{LSMUV}). We will consider the system at some finite temperature $T$, meaning that from here on we will work in Euclidean space with a compactified time direction corresponding to the inverse temperature. That is, all momentum integrals in the timelike direction become a finite-temperature Matsubara sum.

\section{FRG approach}
\label{sec:FlowEquation}

\subsection{Basics}
\label{sec:Basics}

In this section, we introduce the concept of the FRG method and derive flow equations for the couplings of the LSM to incorporate quantum and thermal fluctuations. We do not intend to give a comprehensive description on the technique; the reader is referred to the following reviews \cite{Berges:2000ew,Dupuis:2020fhh}.

The core of the FRG formalism lies on the scale-dependent quantum effective action, $\Gamma_k$, which, unlike the ordinary quantum effective action $\Gamma$, incorporates fluctuations only with momenta $q \gtrsim k$, where $k$ plays the role of a scale separation variable. This construction is achieved by introducing a momentum-dependent mass term into the classical action, usually called the regulator, which gives a large mass to the IR modes, effectively freezing and, thus, excluding them from the path integral. As $k\rightarrow 0$, one demands that the regulator vanishes, leading to $\Gamma_{k=0}=\Gamma$, while for $k\rightarrow \infty$ (or practically some UV cutoff $\Lambda_{\rm UV}$) requires that the regulator tends to infinity for all modes, realizing that $\Gamma_{k\rightarrow \infty}=S$, $S$ being the classical action, upon which our quantum field theory is built.

The scale-dependent effective action, $\Gamma_k$ obeys the Wetterich equation \cite{Wetterich:1992yh}:
\begin{eqnarray}
\label{Wet1}
\partial_k \Gamma_k = \frac12 {\rm Tr} \Big\{ \partial_k R_k [\Gamma_k''+R_k]^{-1}\Big\},
\end{eqnarray}
where $\Gamma_k''$ is the second functional derivative of $\Gamma_k$ with respect to the fields, and the trace operation needs to be taken both in the matrix and in the functional sense. By introducing the differential operator
\begin{eqnarray}
\tilde{\partial}_k = \partial_k R_k \frac{\partial}{\partial R_k},
\end{eqnarray}
Eq.~(\ref{Wet1}) can be conveniently written as
\begin{eqnarray}
\label{Wet1b}
\partial_k \Gamma_k = \frac12 \tilde{\partial}_k {\rm Tr}\, {\rm Log} \, [\Gamma_k''+R_k], \label{FlowEq2}
\end{eqnarray}
where once again, both the Tr and Log operations need to be taken in the functional and matrix senses. One may also say that $\tilde{\partial}_k$ is a differential operator that acts on only the $k$ dependence of $R_k$.

The right-hand side (rhs) of Eq.~(\ref{Wet1}) or (\ref{Wet1b}) is generally nonlocal since $\Gamma_k''$ remains as a function of ${\cal S}^a(x)$ and ${\cal P}^a(x)$. In the present analysis, however, we employ the local potential approximation (LPA) and ignore nonlocalities. In the LPA, the scale-dependent effective action is approximated as ($\int_x \equiv \int d^4x_E$)
\begin{eqnarray}
\label{LPA}
\Gamma_k = \int_x \Big[ {\rm tr}[\partial_i\Sigma^\dagger\partial_i\Sigma] + \bar{c}{\rm tr}[\zeta^\dagger\Sigma + \Sigma^\dagger\zeta] + V_k \Big],
\end{eqnarray}
where $V_k$ is now a local function, called the scale-dependent effective potential. Note that, we have adopted the Euclidean signature compared to Eq.~(\ref{LSMUV}). Also note tha the explicit symmetry-breaking term does not depend on the separation scale $k$, which follows from the fact that couplings corresponding to linear terms in the field variable do not renormalize. By employing the LPA ansatz (\ref{LPA}), what we need to establish is a flow equation for $V_k$. 

Using Eq.~(\ref{LPA}), the second derivative of $\Gamma_k$ with respect to the fields in Fourier space reads
 \begin{eqnarray}
 \label{gamma2}
 \Gamma_k''(p,q)  = \big( p^2 + V_k'' \big)\delta(p+q),
 \end{eqnarray}
where the momentum variables contain both the frequencies and the three-momenta: therefore, in accordance, $p^2=\omega_n^2+{\bm p}^2$ with $\omega_n=2\pi nT$ ($n\in {\mathbb Z}$) being the Matsubara frequencies. In the present analysis, we adopt the so-called 3D Litim regulator \cite{Litim:2001up}, which, in Fourier space, depends only on the three-momentum:
 \begin{eqnarray}
 R_k(p) = (k^2-{\bm p}^2)\Theta(k^2-{\bm p}^2),
 \end{eqnarray}
and it is considered diagonal in flavor space. Plugging Eq.~(\ref{gamma2}) into Eq.~(\ref{FlowEq2}), we can perform the functional trace operation to arrive at
 \begin{eqnarray}
 \label{Wet2}
 \partial_k V_k = \frac{T}{2} \sum_{n} \int \frac{d^3p}{(2\pi)^3} \tilde{\partial}_k {\rm tr}\, {\rm \log}\, [\omega_n^2+{\bm p}^2+V_k''+R_k({\bm p})],\nonumber\\
 \end{eqnarray}
where now both the trace and log operations need to be considered only in the matrix sense, and the rhs includes one single-momentum integral only. Note that, owing to the LPA, wave function renormalizations of all field variables are omitted. The summation over $n$ stands for the infinite sum of the Matsubara frequencies $\omega_n$, explicitly showing that we are working at some finite temperature $T$. At the UV scale $k=\Lambda_{\rm UV}$, and $V_{k=\Lambda_{\rm UV}}$ is reduced to the classical potential (\ref{VUV}), while $V_{k=0}$ becomes the full quantum effective potential. In solving the flow equation, we need to integrate from $k=\Lambda_{\rm UV}$ to $k=0$ using the initial condition dictated by Eq.~(\ref{VUV}).

\subsection{General structures}
\label{sec:Ansatz}

Since chiral symmetry is realized linearly in our framework, the effective action (and, thus, the effective potential) inherits the symmetry properties of the classical action. That is, $V_k$ can depend on only independent chiral invariants, $V_k = V_k(I_1,I_2,I_A)$. Within the present LSM framework, the spontaneous breaking of chiral symmetry is triggered by the appearance of a nonzero ${\cal S}^0$ mean field, $\sigma_0 \equiv \langle {\cal S}^0\rangle$. In this vacuum configuration the $I_2$ invariant vanishes: $\langle I_2\rangle=0$. After taking into account quantum corrections to the effective potential, higher orders of $I_2$, i.e., $I_2^2$, $I_2^3$,..., are expected to be generated. These are, however, nonrenormalizable interactions, which are expected to produce only small contributions during the flows. For this reason, we employ the following ansatz for the potential at the quantum level:
\begin{eqnarray}
V_k(I_1,I_2,I_A) \approx U_k(I_1,I_A) + C_k(I_1,I_A)I_2. \label{VkAnsatz}
\end{eqnarray}
Using Eq.~(\ref{Wet2}) one, in principle, can derive functional flow equations for the $U_k$ and $C_k$ functions. In this exploratory study, however, we restrict ourselves for the first few Taylor coefficients and approximate $U_k$ and $C_k$ as
\begin{eqnarray}
U_k &\approx& m_k^2I_1 + \lambda_{1,k}I_1^2 + a_k I_A + c_{1,k} I_A^2 + c_{2,k} I_1 I_A, \nonumber\\
C_k &\approx& \lambda_{2,k},
\label{VkAssume}
\end{eqnarray}
where $m_k^2$, $\lambda_{1,k}$, $a_k$, $c_{1,k}$, $c_{2,k}$ and $\lambda_{2,k}$ are field-independent flowing coupling constants. Note that the structure of Eq.~(\ref{VkAssume}) is unchanged from the classical one (\ref{Vcl}), meaning that only renormalizable operators are taken into account. Each coupling explicitly contains a subscript $k$ stressing that they are defined at scale $k$.

Introducing the notation $(\Gamma_k'')^{-1}(p,q)={\cal D}_k(p)\delta(q+p)$, from Eq.~(\ref{gamma2}) we have
\begin{eqnarray}
({\cal D}_k^{-1})^{AB}(p) = \big(\omega_n^2+{\bm p}^2+R_k({\bm p})\big)\delta^{AB}+(V_k'')^{AB}\ , \label{DkInv}
\end{eqnarray}
with
\begin{eqnarray}
(V_k'')^{AB} = \left(
\begin{array}{cc}
\frac{\partial^2 V_k}{\partial {\cal S}^a \partial{\cal S}^b} & \frac{\partial^2 V_k}{\partial {\cal S}^a \partial{\cal P}^b}  \\
\frac{\partial^2 V_k}{\partial {\cal P}^a \partial{\cal S}^b} & \frac{\partial^2 V_k}{\partial {\cal P}^a \partial{\cal P}^b} \\ 
\end{array}
\right)\ . \label{VkDerivative}
\end{eqnarray}
We note that both $({\cal D}_k^{-1})^{AB}$ and $(V_k'')^{AB}$ are $12\times12$ matrices. since in total $12$ hadrons are considered. Using the above notation, from Eq.~(\ref{Wet2}) we have
\begin{eqnarray}
\label{Wet2b}
\partial_k V_k = \frac{T}{2}\sum_n \int \frac{d^3p}{(2\pi)^3}\tilde{\partial}_k {\rm tr}{\rm log} ({\cal D}^{-1}_k) (p).
\end{eqnarray}

\subsection{Flowing couplings}
\label{sec:EffectiveAction}

Flow equations for the coupling constants, defined in Eq.~(\ref{VkAssume}), can be derived by expanding both sides of Eq.~(\ref{Wet2b}) in terms of the invariants $I_1$, $I_A$ and $I_2$. To this end, we expand the rhs of Eq.~(\ref{Wet2b}) up to ${\cal O}(\Sigma^4)$ as

\begin{eqnarray}
\partial_k V_k &=& \frac{T}{2}\sum_n\int\frac{d^3p}{(2\pi)^3}\tilde{\partial}_k{\rm tr}\, {\rm ln}\Big[{\cal D}_{0,k}^{-1}+V_{(2)k}''\Big] \nonumber\\
&\approx& \frac{T}{2}\sum_n\int\frac{d^3p}{(2\pi)^3}\tilde{\partial}_k{\rm tr}\Big[{\cal D}_{0,k}^{-1}\Big] \nonumber\\
&&+  \frac{T}{2}\sum_n\int\frac{d^3p}{(2\pi)^3}\tilde{\partial}_k{\rm tr}\Big[{\cal D}_{0,k}V_{(2)k}''\Big] \nonumber\\
&& - \frac{T}{4}\sum_n\int\frac{d^3p}{(2\pi)^3}\tilde{\partial}_k{\rm tr}\Big[{\cal D}_{0,k}V_{(2)k}''{\cal D}_{0,k}V_{(2)k}''\Big], \nonumber\\ \label{OmegaExpanded}
\end{eqnarray}
where we have separated $V_k''$ into ${\cal O}(\Sigma^0)$ and ${\cal O}(\Sigma^2)$ contributions, denoted by $V_{(0)k}''$ and $V_{(2)k}''$, respectively: $V_k''=V_{(0)k}''+V_{(2)k}''$, and the inverse of the propagator matrix is also divided as ${\cal D}_k^{-1} = {\cal D}_{0,k}^{-1} + V_{(2)k}''$, where 
\begin{eqnarray}
{\cal D}_{0,k}^{-1} &=& \omega_n^2+{\bm p}^2+R_k({\bm p})+V_{(0)k}'' \ . \label{D0kInverse}
\end{eqnarray}
The first term in Eq.~(\ref{OmegaExpanded}) corresponds to the vacuum energy, which has nothing to do with the flow equations of the couplings. It is the second and third terms which determine the latter in the end.

Here, as a simple demonstration we focus on
\begin{eqnarray}
\partial_k V_k^{(2)} \equiv \frac{T}{2}\sum_n\int\frac{d^3p}{(2\pi)^3}\tilde{\partial}_k{\rm tr}\Big[{\cal D}_{0,k}V_{(2)k}''\Big] \ , \label{Vk2Def}
\end{eqnarray}
which leads to flow equations for $m_k$ and $a_k$. The trace of Eq.~(\ref{Vk2Def}) is readily performed, since only one ${\cal D}_{0,k}$ is included. With the help of the Matsubara-summation formula
\begin{eqnarray}
T\sum_n \frac{1}{\omega_n^2+\epsilon^2} =\frac{1}{2\epsilon}\big[1+2 f_B(\epsilon)\big]\ ,
\end{eqnarray}
where $f_B(\epsilon)=1/({\rm e}^{\epsilon/T}-1)$ is the Bose-Einstein distribution function, $\partial_k V_k^{(2)}$ is expressed as
\begin{eqnarray}
\partial_k V_k^{(2)} = \partial_k V_{M,k}^{(2)} + \partial_k V_{B,k}^{(2)}\ , \label{Vk2MAndB}
\end{eqnarray}
where
\begin{eqnarray}
\partial_k V_{M,k}^{(2)} &=& \frac{1}{2}\int\frac{d^3p}{(2\pi)^3} \tilde{\partial}_k \nonumber\\
&& \Bigg(\frac{\sum_{a=1}^3V_{k,{\cal P}^a{\cal P}^a}''^{(2)}+V_{k,{\cal S}^0{\cal S}^0}''^{(2)}}{2E_{k}^+}\Big[1 + 2f_B(E_{k}^+)\Big] \nonumber\\
&& + \frac{\sum_{a=1}^3V_{k,{\cal S}^a{\cal S}^a}''^{(2)}+V_{k,{\cal P}^0{\cal P}^0}''^{(2)}}{2E_{k}^-}\Big[1 + 2f_B(E_{k}^-)\Big] \Bigg) \, \nonumber\\
\end{eqnarray}
and
\begin{eqnarray}
\partial_k V_{B,k}^{(2)}&=& \frac{1}{2}\int\frac{d^3p}{(2\pi)^3}\tilde{\partial}_k \nonumber\\
&&\Bigg(\frac{V_{k,{\cal P}^4{\cal P}^4}''^{(2)}+V_{k,{\cal P}^5{\cal P}^5}''^{(2)}}{2E_{k}^+}\Big[1 + 2f_B(E_{k}^+) \Big] \nonumber\\
&&+ \frac{V_{k,{\cal S}^4{\cal S}^4}''^{(2)}+V_{k,{\cal S}^5{\cal S}^5}''^{(2)}}{2E_{k}^-}\Big[1 + 2f_B(E_{k}^-)\Big]  \Bigg), \nonumber\\
\end{eqnarray}
representing contributions that arise from mesonic and baryonic fluctuations, respectively. In these expressions, $E_k^\pm = \sqrt{{\bm p}^2+R_k({\bm p})+(m^\pm_{{\rm eff},k})^2}$ are the dispersion relations of the fluctuations, where the $k$-dependent effective masses $m^\pm_{{\rm eff},k}$ read
\begin{eqnarray}
(m^\pm_{{\rm eff},k})^2 = m_k^2\pm a_k \ . \label{KEffMass}
\end{eqnarray}
We note that $E_k^+$ corresponds to $\sigma$, $\pi$, and $B$ ($\bar{B}$), while $E_k^-$ for $\eta$, $a_0$,  and $B'$ ($\bar{B}'$). 

Since $\partial_k V_{M,k}^{(2)} $ and $\partial_k V_{B,k}^{(2)}$ are proportional to quadratic forms in terms of ${\cal S}^a$ and ${\cal P}^a$, due to symmetry their sum must combine into contributions proportional to $I_1$ and $I_A$ appropriately, i.e., 
\begin{eqnarray}
\label{separ}
\partial_k V_{M,k}^{(2)} &=& \partial_k \hat{V}^{I_1}_{M,k}I_1+\partial_k \hat{V}^{I_A}_{M,k}I_A + \partial_k\delta V_{45,k}\ , \nonumber\\
\partial_k V_{B,k}^{(2)} &=& \partial_k \hat{V}^{I_1}_{B,k}I_1+\partial_k \hat{V}^{I_A}_{B,k}I_A  - \partial_k\delta V_{45,k}\ ,
\end{eqnarray}
where $\delta V_{45,k}$ is a function of $P^4$, $P^5$, $S^4$ and $S^5$. This $\delta V_{45,k}$ appears to ensure the $SU(4)$-symmetry property, which relates mesons and baryons, and it drops out from $\partial_k V_k^{(2)}$ in Eq.~(\ref{Vk2MAndB}). Here, we only stress that when considering mesonic and baryonic contributions separately in Sec. V, we will omit $\delta V_{45,k}$. Inserting (\ref{separ}) into (\ref{Vk2MAndB}) yields
\begin{eqnarray}
\partial_k V_k^{(2)} &=& \left( \partial_k \hat{V}^{I_1}_{M,k}+\partial_k \hat{V}^{I_1}_{B,k} \right) I_1 \nonumber\\
&&+ \left( \partial_k \hat{V}^{I_A}_{M,k}+\partial_k \hat{V}^{I_A}_{B,k} \right) I_A \ ,
\end{eqnarray}
where
\begin{widetext}
\begin{eqnarray}
\partial_k \hat{V}^{I_1}_{M,k} &=& \frac{1}{2}\int\frac{d^3p}{(2\pi)^3}\tilde{\partial}_k\Bigg(\frac{4c_{1,k}+12c_{2,k}+20\lambda_{1,k}+3\lambda_{2,k}}{4E^+_{k}} \Big[1 + 2f_B(E_{k}^+)\Big] \nonumber\\
&&\hspace{2.1cm}+\frac{4c_{1,k}-12c_{2,k}+20\lambda_{1,k}+3\lambda_{2,k}}{4E^-_{k}}\Big[1 + 2f_B(E_{k}^-)\Big]\Bigg) \ , \label{VI1M} \\
\partial_k \hat{V}^{I_1}_{B,k} &=& \frac{1}{2}\int\frac{d^3p}{(2\pi)^3}\tilde{\partial}_k \Bigg( \frac{4c_{2,k}+8\lambda_{1,k}+2\lambda_{2,k}}{4E_{k}^+}\Big[1 + 2f_B(E_{k}^+) \Big] + \frac{-4c_{2,k}+8\lambda_{1,k}+2\lambda_{2,k}}{4E_{k}^-}\Big[1 + 2f_B(E_{k}^-) \Big] 
 \Bigg)\ ,
 \end{eqnarray}
\begin{eqnarray}
\partial_k \hat{V}^{I_A}_{M,k} &=& \frac{1}{2}\int\frac{d^3p}{(2\pi)^3}\tilde{\partial}_k \Bigg(\frac{20c_{1,k}+12c_{2,k}+4\lambda_{1,k}-3\lambda_{2,k}}{4E_{k}^+}\Big[1 + 2f_B(E_{k}^+)\Big] \nonumber\\
&&\hspace{1.8cm}+\frac{-20c_{1,k}+12c_{2,k}-4\lambda_{1,k}+3\lambda_{2,k}}{4E_{k}^-} \Big[1 + 2f_B(E_{k}^-)\Big]\Bigg) \ , 
\end{eqnarray}
and
\begin{eqnarray}
\partial_k \hat{V}^{I_A}_{B,k} = \frac{1}{2}\int\frac{d^3p}{(2\pi)^3}\tilde{\partial}_k\Bigg( \frac{8c_{1,k}+4c_{2,k}-2\lambda_{2,k}}{4E_{k}^+}\Big[1 + 2f_B(E_{k}^+)  \Big] + \frac{-8c_{1,k}+4c_{2,k}+2\lambda_{2,k}}{4E_{k}^-}\Big[1 + 2f_B(E_{k}^-)\Big]\Bigg)\ . \label{VIAB}
\end{eqnarray}
\end{widetext}
Therefore, the flow equations for $m^2_k$ and $a_k$ can be easily obtained as
\begin{eqnarray}
\partial_k m^2_k &=& \partial_k\hat{V}^{I_1}_{M,k} + \partial_k\hat{V}^{I_1}_{B,k}\ , \nonumber\\
\partial_k a_k &=& \partial_k\hat{V}^{I_A}_{M,k} + \partial_k\hat{V}^{I_A}_{B,k}\ , \label{MkAkFlow}
\end{eqnarray}
and we note once again that the $\tilde{\partial}_k$ operator acts only on $R_k$ in Eqs.~(\ref{VI1M})--(\ref{VIAB}). We stress that, in Eq.~(\ref{MkAkFlow}), mesonic and baryonic contributions are intentionally separated, as it turns out to be useful for later analysis. The $\tilde{\partial}_k$ operation in the rhs is easily handled by means of Eq.~(\ref{KDFormula}). Note that, since $\partial_k R_k({\bm p}) \sim \Theta(k^2-{\bm p}^2)$, all momentum integrals are restricted to $|{\bm p}| \in [0,k]$, leading to ${\bm p}^2 +R_k({\bm p}) \rightarrow k^2$ in the integrands. As a result, all integrals become trivial, and one formally substitutes $\int d^3p/(2\pi)^3 \rightarrow k^3/6\pi^2$.

Unlike demonstrated above, the third term in the rhs of Eq.~(\ref{OmegaExpanded}) contains three types of contributions, i.e., purely mesonic ones, purely baryonic ones, and those that mix the two. In accordance, in the forthcoming analysis, we define the first one as the purely mesonic contribution, while the second and third ones will belong to the contributions classified as baryonic. This separation enables us to obtain flow equations for $\lambda_{1,k}$, $\lambda_{2,k}$, $c_{1,k}$, and $c_{2,k}$ as a sum of mesonic and baryonic fluctuations. We do not go through the explicit calculations for the aforementioned flows; they are summarized in Appendix~\ref{sec:kTildeDerivative}.

Before closing this subsection, we comment on the anomalous structure of the effective potential. When the $U(1)_A$ anomaly is switched off, $a_k=0=c_{1,k}=c_{2,k}$ holds. In this case, the $U(1)$ axial charges are conserved, or, intuitively speaking, chirality-flipping processes of a single quark do not occur. As a result, all fluctuations universally obey the dispersion relation of $E_k^+=E_k^-=\sqrt{{\bm p}^2+R_k({\bm p})+m_k^2}$. Consequently, all the anomalous components of the potential vanish: $\hat{V}^{I_A}_{M,k}=\hat{V}^{I_A}_{B,k}=\hat{V}_{M,k}^{I_A^2}=\hat{V}_{B,k}^{I_A^2}=\hat{V}_{M,k}^{I_1I_A}=\hat{V}_{M,k}^{I_1I_A}=0$.

\begin{table}[t]
\begin{center}
  \begin{tabular}{c|cccccc}  \hline\hline
$M_\eta$ [GeV] & $m_\Lambda^2$ [GeV$^2$] & $a_\Lambda$ [GeV] & $\lambda_{1,\Lambda}$ & $\lambda_{2,\Lambda}$ & $c_{1,\Lambda}$ & $c_{2,\Lambda}$ \\ \hline
$0.3$ & $-0.0426$ & $-0.032$ & $1.25$ & $5$ & $0$ & $0$ \\
$0.5$ & $0.0465$ & $-0.108$ & $1.25$ & $5$ & $0$ & $0$ \\
$0.95$ & $0.385$ & $-0.428$ & $1.25$ & $5$ & $0$ & $0$ \\
\hline \hline
 \end{tabular}
\caption{The parameters at $k=\Lambda_{\rm UV}$ for $\mathring{M}_\eta=0.3$, $0.5$, and $0.95$ GeV with $\Lambda_{\rm UV}=1$ GeV. }
\label{tab:Input}
\end{center}
\end{table}

\section{Numerical results}
\label{sec:Result}

In this section, we proceed with numerical analyses by solving the flow equations derived in Sec.~\ref{sec:EffectiveAction}. 

First, in Sec.~\ref{sec:Inputs}, our criterion to fix initial conditions for the flow equations is explained. Then, in Secs.~\ref{sec:HadronMass} and Sec.~\ref{sec:AnomalyT}, we present our main results on the hadron masses and modifications of the anomaly coefficients at finite temperature. Finally, in Sec.~\ref{sec:TopSusceptibility}, numerical results on the topological susceptibility are shown.

\begin{figure*}[t]
\centering
\hspace*{-0.5cm} 
\includegraphics*[scale=0.45]{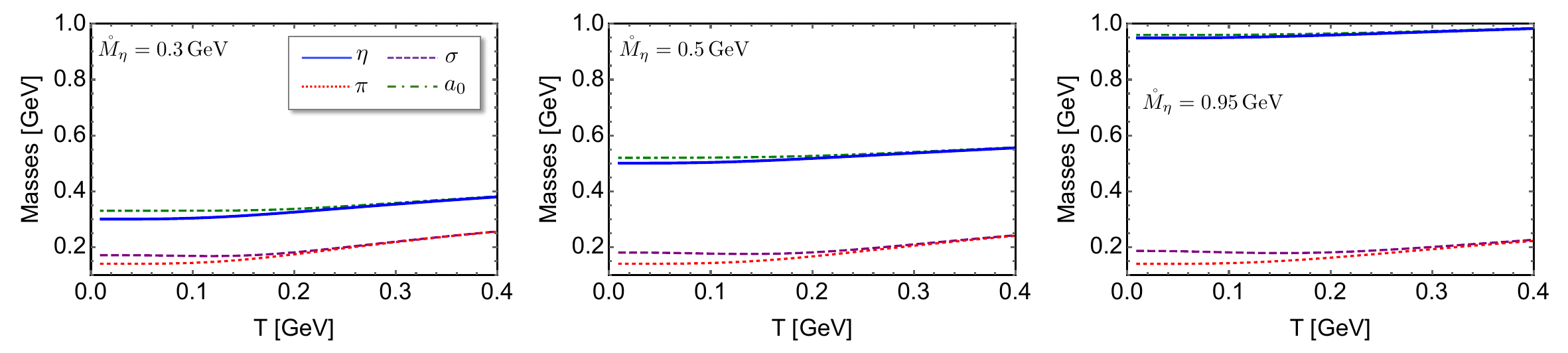}
\caption{$T$ dependencies of the masses of $\eta$, $\pi$, $\sigma$, and $a_0$ mesons with $\mathring{M}_\eta=0.3$, $0.5$, and $0.95$ GeV. }
\label{fig:HadronMass}
\end{figure*}

\begin{figure*}[t]
\centering
\hspace*{-0.5cm} 
\includegraphics*[scale=0.45]{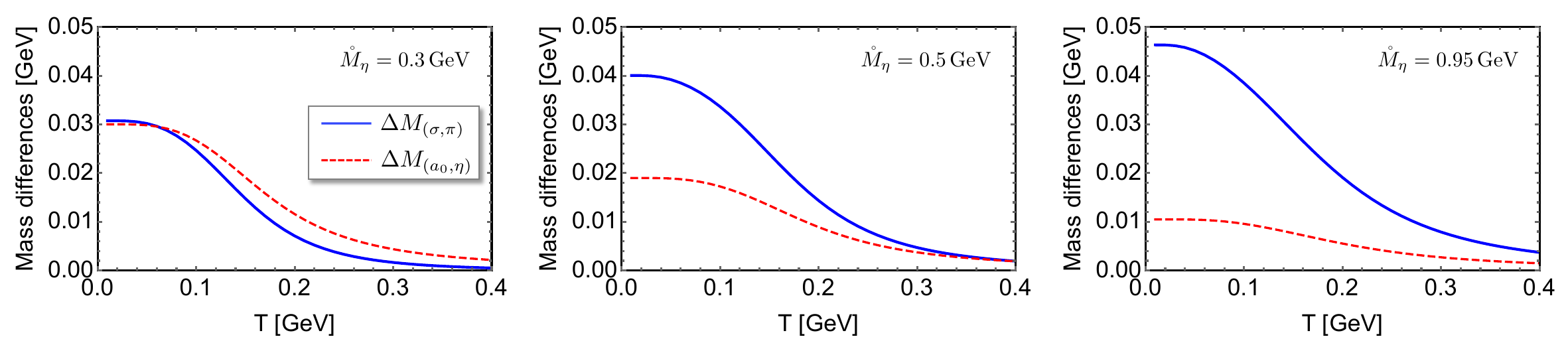}
\caption{$T$ dependencies of the mass differences $\Delta M_{(\sigma.\pi)}$ and $\Delta M_{(a_0.\eta)}$ with $\mathring{M}_\eta=0.3$, $0.5$, and $0.95$ GeV. }
\label{fig:MassDiff}
\end{figure*}

\begin{figure*}[t]
\centering
\hspace*{-0.5cm} 
\includegraphics*[scale=0.45]{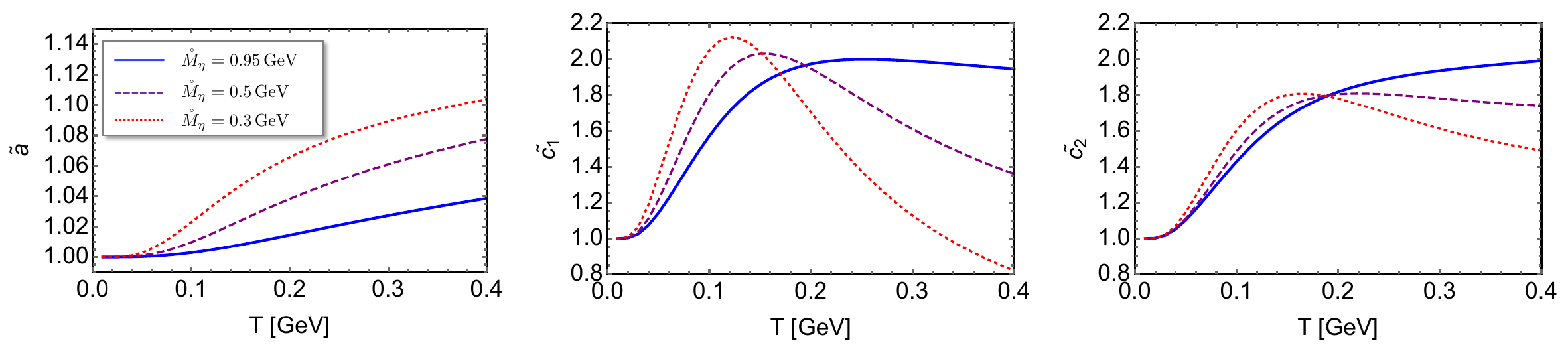}
\caption{$T$ dependencies of the normalized anomaly coefficients $\tilde{a}$, $\tilde{c}_1$, and $\tilde{c}_2$ defined in Eq.~(\ref{AnomalyNormalized}), with $\mathring{M}_\eta=0.3$, $0.5$ and $0.95$ GeV. }
\label{fig:Anomaly}
\end{figure*}

\subsection{Inputs}
\label{sec:Inputs}

We have six parameters in the potential~(\ref{VUV}), which also correspond to initial conditions of the flow equations. In the present analysis, we assume the large $N_c$ ansatz for the potential for the UV potential, where the double-trace terms and the anomalous four-point interactions are suppressed~\cite{Witten:1979kh,DiVecchia:1980yfw}, leading to $\lambda_{2,\Lambda}=4\lambda_{1,\Lambda}$ [see Eq.~(\ref{Lambda1Lambda2})] and $c_{1,\Lambda}=c_{2,\Lambda}=0$. In addition, it is found in the numerics that $\lambda_{2,\Lambda}$ must satisfy $\lambda_{2,\Lambda}\lesssim5$ to produce a reasonable reduction of the chiral condensate at finite temperature.\footnote{When choosing $\lambda_2\gtrsim 5$, the chiral condensate can increase at finite $T$, which is obviously unphysical.} Hence, we adopt $\lambda_{2,\Lambda}=5$ and accordingly $\lambda_{1,\Lambda}=1.25$ as an adequate demonstration. The quadratic couplings at at the UV scale, $m_{\Lambda}^2$ and $a_\Lambda$, are then fixed by fitting the pion and the $\eta$ meson masses in the vacuum. The former will be fixed to $\mathring{M}_\pi=0.14$ GeV, while for the latter we investigate three different scenarios, $\mathring{M}_\eta=0.3, 0.5, 0.95$ GeV.\footnote{The symbol $\mathring{M}$ stands for the corresponding vacuum value. Also, in what follows, we will omit the subscript $0$ representing $k=0$ in the masses and parameters, as a shorthand notation.} The reason is that our main motivation in this study is to explore the fate of $U(1)_A$ anomaly at finite temperature, and one might be curious about such dependencies as a function of the anomaly strength itself. All the masses are evaluated at $T=0$, after integrating down the flow equations to $k=0$: see their respective expressions, (\ref{PiMass}). Finally, the explicit breaking parameter $\bar{c}m_q$ in Eq.~(\ref{Lex}) is determined by the decay constant $f_{\pi}=0.093$ GeV, which equals to the vacuum expectation value $\mathring{\sigma}_0$ due to the corresponding Ward identity. In our study, the initial UV scale is set to be $\Lambda_{\rm UV} = 1$ GeV, and all the numerical values of the determined parameters are summarized in Table~\ref{tab:Input}. We also note that current lattice simulations in QC$_2$D are being carried out with a heavy pion mass, e.g., $\mathring{M}_\pi^{\rm lattice}\sim 0.7$ GeV~\cite{Boz:2019enj,Astrakhantsev:2020tdl,Iida:2024irv}. Hence, predictions of our present work may also be considered as an important benchmark for future lattice computations with lighter pion masses.

\subsection{Hadron masses at finite $T$}
\label{sec:HadronMass}

In this subsection, we examine $T$ dependencies of the hadron masses by solving the flow equations. Their analytic mass formulas are summarized in Appendix~\ref{sec:HadronMassFormula}, which indicate that $M_{\pi}=M_{B(\bar{B})}$ while $M_{a_0}=M_{B'(\bar{B}')}$ at any $T$, stemming from the pseudoreality ${\bm 2}\simeq \bar{\bm 2}$ of QC$_2$D~\cite{Suenaga:2022uqn,Suenaga:2025sln}.

Figure ~\ref{fig:HadronMass} shows the resultant hadron masses for $\mathring{M}_\eta=0.3$, $0.5$, and $0.95$ GeV at finite $T$. All the masses are mostly dominated by the quadratic terms $m^2$ and $a$, since the quartic couplings are suppressed; particularly, $\lambda_2$ (and, accordingly, $\lambda_1=\lambda_2/4$) are found to remain small throughout the flow. Hence, from the mass formulas of the hadrons provided in Eqs.~(\ref{PiMass}) and~(\ref{SigmaMass}), mass differences between $\sigma$ and $\pi$ or between $a_0$ and $\eta$ are always small as $M^2_\sigma, M^2_\pi \sim m^2+ a$ while $M^2_{a_0} , M^2_\eta \sim m^2-a$. We note that this approximation applies to any $k$ throughout the flow.

\begin{figure}[t]
\centering
\hspace*{-0.5cm} 
\includegraphics*[scale=0.6]{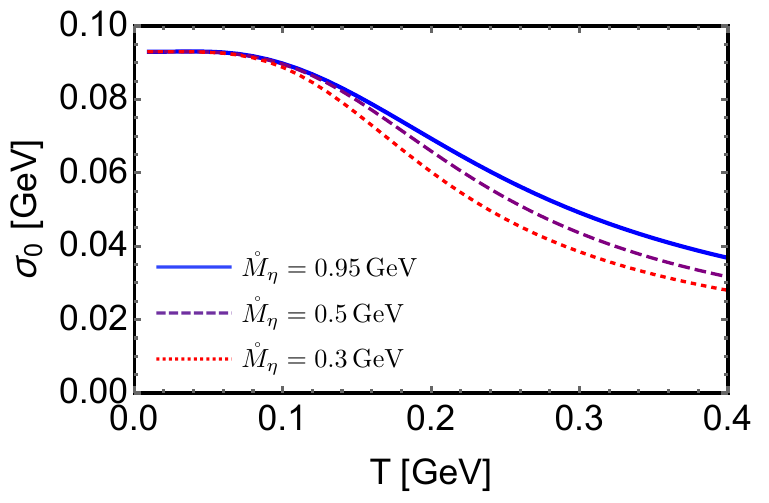}
\caption{$T$ dependencies of the mean field $\sigma_0$ with $\mathring{M}_\eta=0.3$, $0.5$, and $0.95$ GeV. }
\label{fig:Chiral}
\end{figure}

Within the present approximation, the strength of thermal excitations is determined by the size of the effective masses~(\ref{KEffMass}); smaller values of $m_{{\rm eff},k}^{\pm}$ throughout the flows yield more thermal effects in the end. Thus, from the above argument one can understand that a smaller value of $\mathring{M}_\eta$ results in more significant thermal contributions to the potential. This property is also reflected by the reduction rate of the chiral condensate, as shown in Fig.~\ref{fig:Chiral}. In fact, this figure indicates that the smaller $\mathring{M}_\eta$ we take, the more rapidly the chiral condensate drops. The pseudocritical temperature read off from Fig.~\ref{fig:Chiral} is $T_{\rm pc} \gtrsim 0.25$ GeV, which is higher than the one for $N_c=3$, where $T_{\rm pc}$ is estimated to be $T_{\rm pc}\sim0.15$ GeV. The slightly higher $T_{\rm pc}$ is also our prediction for QC$_2$D with a light pion mass.

Figure~\ref{fig:HadronMass} indicates that the $\sigma$ meson mass is always $\mathring{M}_\sigma\sim0.18$ GeV at vanishing temperature, which is smaller than the two-pion threshold $2\mathring{M}_\pi = 0.28$ GeV. That is, our current analysis predicts a stable $\sigma$ meson in QC$_2$D with the physical pion mass, unlike the $N_c=3$ world. It would be interesting to see whether future lattice simulations can confirm such a light $\sigma$ meson.

In order to take a closer look at mass degeneracies of the chiral partners, we depict the $T$ dependencies of 
\begin{eqnarray}
\Delta M_{(\sigma,\pi)} \equiv M_{\sigma}-M_\pi\ , \ \ \Delta M_{(a_0,\eta)} \equiv M_{a_0}-M_{\eta} \ ,
\end{eqnarray}
in Fig.~\ref{fig:MassDiff}. At finite $T$, the mass differences reduce because of chiral symmetry restoration, for which the chiral partner structures are also clearly realized. Similar mass degeneracies were also found at the high-density limit with vanishing temperature~\cite{Suenaga:2022uqn,Suenaga:2023xwa}. We note that $\Delta M_{(\sigma,\pi)}$ and $\Delta M_{(a_0,\eta)}$ are at most of ${\cal O}(10)$ MeV, regardless of the value of $\mathring{M}_\eta$, since the $\eta$ meson mass is predominantly generated by the axial anomaly, as shown in Eq.~(\ref{PiEtaMass}), and the anomaly does not play a significant role in the mass degeneracies of chiral partners. The mass difference $\Delta M_{(\sigma,\pi)}$ gets enhanced for a heavier $\eta$ meson, which mainly results from the increase of $c_2$. We note that, when the anomaly is turned off, the tendency is opposite; the mass difference $\Delta M_{(\sigma,\pi)}$ is reduced for heavier $\mathring{M}_\eta$ when $a=c_1=c_2=0$.

\begin{figure}[t]
\centering
\hspace*{-0.4cm} 
\includegraphics*[scale=0.6]{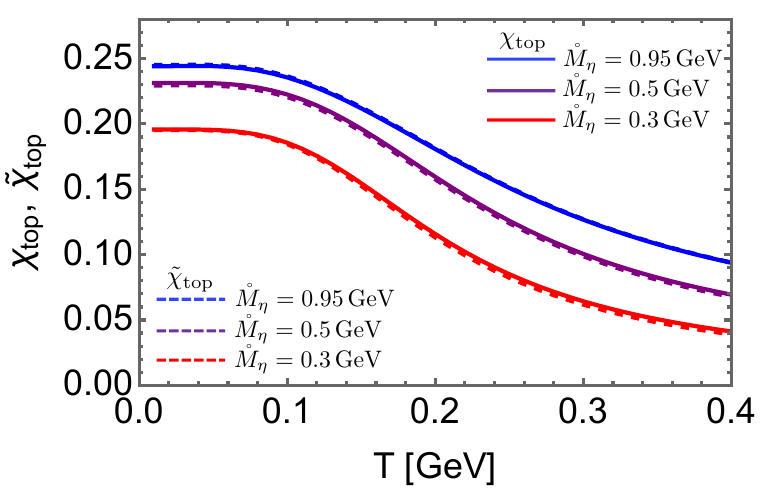}
\caption{$T$ dependencies of the normalized topological susceptibilities $\chi_{\rm top}$ (solid lines) and $\tilde{\chi}_{\rm top}$ (dashed lines) defined in Eqs.~(\ref{ChiTop1}) and~(\ref{ChiTop2}), respectively, with $\mathring{M}_\eta=0.3$, $0.5$, and $0.95$ GeV. }
\label{fig:ChiTop}
\end{figure}

\subsection{Enhancement of anomaly effects at finite $T$}
\label{sec:AnomalyT}

Here we numerically study modifications of the anomaly coefficients $a$, $c_{1}$ and $c_{2}$ at finite temperature. Depicted in Fig.~\ref{fig:Anomaly} are the resultant $T$ dependencies of the coefficients normalized by their respective vacuum values: 
\begin{eqnarray}
\tilde{a} \equiv \frac{a}{a_{T=0}}\ , \ \ \tilde{c}_1 \equiv \frac{c_1}{c_{1,T=0}}\ , \ \ \tilde{c}_2 \equiv \frac{c_2}{c_{2,T=0}} \ . \label{AnomalyNormalized}
\end{eqnarray} For any values of $\mathring{M}_\eta$, all the coefficients exhibit significant increase at finite temperature; that is, the anomaly enhancement in hot medium is predicted to be similar in QC$_2$D to the case of (three-color) QCD~\cite{Fejos:2016hbp}. 

The enhancement of the quadratic anomaly coefficient $a$ is milder than the quartic ones, $c_1$ and $c_2$, where a larger $\mathring{M}_\eta$ leads to a smaller increase. For $c_1$ and $c_2$, the same ordering of the enhancement is seen up to $T\sim0.2$ GeV, while at higher temperatures it is flipped. In general, the coefficients show a significant increase at low $T$, while at high $T$ the enhancements become tempered and begin to decrease. Note that the inversions occur when the chiral condensate starts to reduce (see Fig.~\ref{fig:Chiral}). In addition, such nonmonotonic behaviors appear at lower $T$ for smaller values of $\mathring{M}_\eta$. We also note that $a<0$, $c_1<0$, and $c_2>0$ at any $T$.

\subsection{Topological susceptibility}
\label{sec:TopSusceptibility}

Here we present numerical results on the topological susceptibility at finite $T$, which is regarded a useful observable to study the $U(1)_A$ anomaly effects.

The topological susceptibility $\bar{\chi}_{\rm top}$ is defined as a spacetime-integrated correlation function of a gluon topological operator $Q=(g_s^2/64\pi^2)\epsilon^{\mu\nu\rho\sigma}G_{\mu\nu}^a G^a_{\rho\sigma}$, which reads
\begin{eqnarray}
\bar{\chi}_{\rm top} \equiv -i\int d^4x\langle0|{\rm T} Q(x)Q(0)|0\rangle\ .
\end{eqnarray}
In terms of the fermionic language, this quantity can be reexpressed through two-point functions of pseudoscalar operators $\bar{\psi}i\gamma_5\psi$ and $\bar{\psi}i\gamma_5\tau_f^a\psi$, by the virtue of axial Ward-Takahashi identities~\cite{GomezNicola:2016ssy,Kawaguchi:2020qvg}. Then, within the LSM, imposing matching conditions appropriately, the topological susceptibility is evaluated as (for the detailed derivation, see Ref.~\cite{Kawaguchi:2023olk})
\begin{eqnarray}
\label{chitop}
\bar{\chi}_{\rm top} = i\frac{\mathring{M}_\pi^4\mathring{\sigma}_0^2}{4}\big[ D_\pi(0)-D_\eta(0)\big]\ ,
\end{eqnarray}
where
\begin{eqnarray}
D_\pi(p)\delta^{ab} &=& \int d^4x \langle0|{\rm T}\pi^a(x)\pi^b(0)|0\rangle {\rm e}^{ip\cdot x}\ , \nonumber\\
D_\eta(p)  &=& \int d^4x \langle0|{\rm T}\eta(x)\eta(0)|0\rangle {\rm e}^{ip\cdot x}\ ,
\end{eqnarray}
are pion and $\eta$-meson propagators, respectively, in momentum space.

In the present work, $D_\pi$ and $D_\eta$ are always simply expressed by $D_\pi=i/(p^2-M_\pi^2)$ and $D_\eta=i/(p^2-M_\eta^2)$, respectively. Note that their functional forms do not differ even at finite $T$, owing to the absence of diquark condensates~\cite{Kawaguchi:2023olk}. Hence, the susceptibility reads
\begin{eqnarray}
\chi_{\rm top} \equiv \frac{\bar{\chi}_{\rm top}}{\mathring{M}_\pi^2\mathring{\sigma}_0^2} = \frac{\mathring{M}_\pi^2}{4}\left(\frac{1}{M_\pi^2} - \frac{1}{M_\eta^2}\right)\ , \label{ChiTop1}
\end{eqnarray}
where we have defined the dimensionless topological susceptibility $\chi_{\rm top}$, normalized by the constant $\mathring{M}_\pi^2\mathring{\sigma}_0^2$. The mass difference between the $\eta$ meson and the pion is generated solely by anomaly effects [see Eq.~(\ref{PiEtaMass})]; therefore, it is beneficial to define
\begin{eqnarray}
\tilde{\chi}_{\rm top} \equiv \frac{\mathring{M}_\pi^2}{4}\left(\frac{1}{M_\pi^2} - \frac{1}{M_\pi^2 + \delta^2M_{(\eta,\pi)}}\right)\ , \label{ChiTop2}
\end{eqnarray}
with $\delta^2M_{(\eta,\pi)} \equiv \mathring{M}_\eta^2-\mathring{M}_\pi^2$. $\tilde{\chi}_{\rm top}$ can serve as a reference, for which the effects resulting from the enhancement of the anomaly in the medium can be properly quantified.

Depicted in Fig.~\ref{fig:ChiTop} are the resulting $T$ dependencies of the normalized topological susceptibilities $\chi_{\rm top}$ and $\tilde{\chi}_{\rm top}$ with $\mathring{M}_\eta=0.3$, $0.5$, and $0.95$ GeV. Differences between the solid and dashed curves are always tiny, meaning that the anomaly enhancement effects do not generate sizable contributions to the topological susceptibility. Meanwhile, the dependence of the $\eta$ meson mass on the topological susceptibility is manifest only for larger $\mathring{M}_\eta$, which increases the magnitude of $\chi_{\rm top}$. Up to $T\sim0.1$ GeV, the susceptibilities are almost constant for any values of $\mathring{M}_\eta$, while above this temperature they start to reduce. This reduction is understood by the smooth chiral symmetry restoration exhibited in Fig.~\ref{fig:Chiral}, since the $T$ dependence of the susceptibility~(\ref{ChiTop1}) is governed by
\begin{eqnarray}
\bar{\chi}_{\rm top} \sim \frac{\mathring{M}_\pi^2}{4}\frac{1}{M_\pi^2} \propto \sigma_0, \label{ChiTopRed}
\end{eqnarray}
assuming $M_\pi\ll M_\eta$, followed from Eq.~(\ref{PiMassRelation}).

\begin{figure}[t]
\centering
\hspace*{-0.5cm} 
\includegraphics*[scale=0.48]{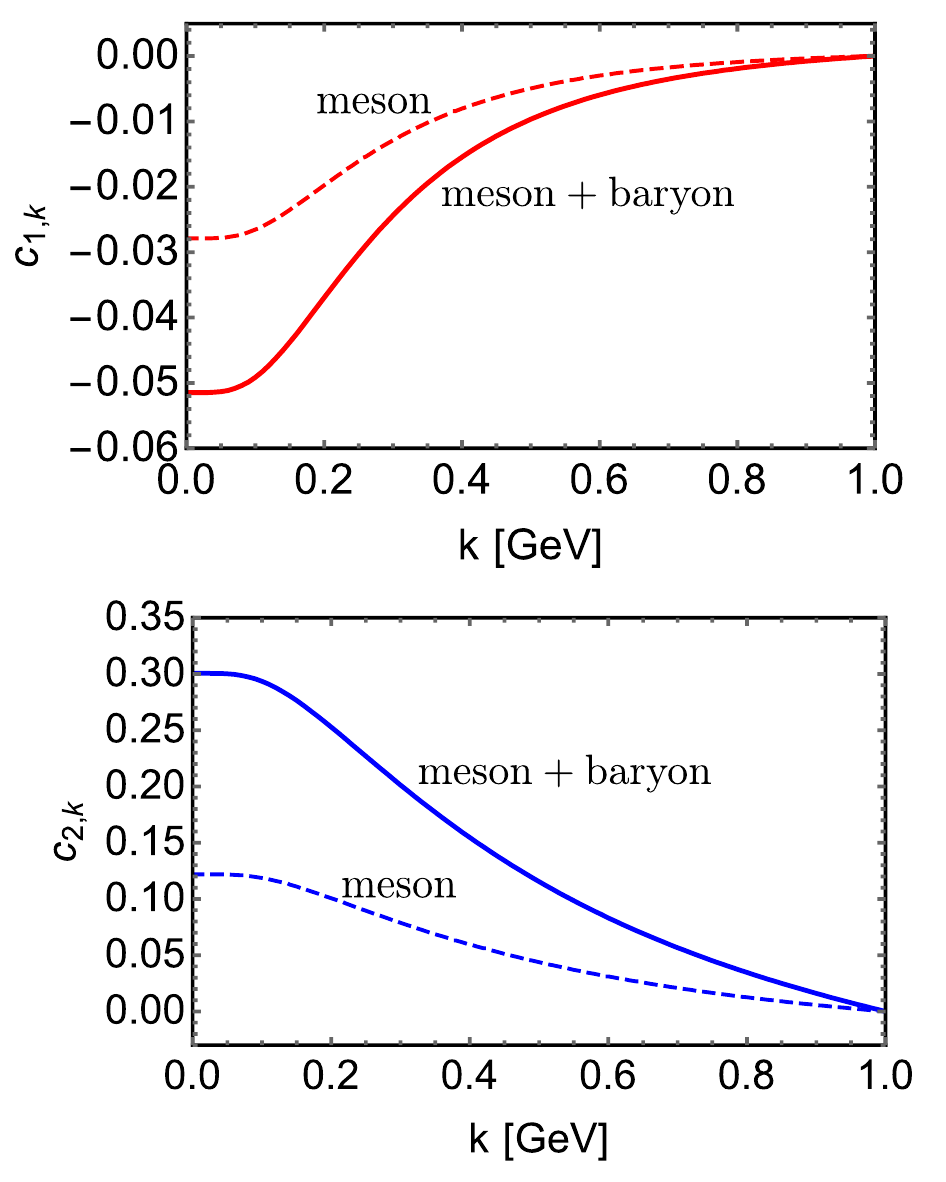}
\caption{$k$ dependencies of the anomaly coefficients $c_{1,k}$ and $c_{2,k}$ evaluated by the full (solid lines) and truncated (dashed lines) analyses at $T=0$, with $\mathring{M}_\eta=0.5$ GeV. Note that the UV scale is set to be $\Lambda_{\rm UV}=1$ GeV.}
\label{fig:AnomalyFlowK}
\end{figure}

\begin{figure}[t]
\centering
\hspace*{-0.5cm} 
\includegraphics*[scale=0.5]{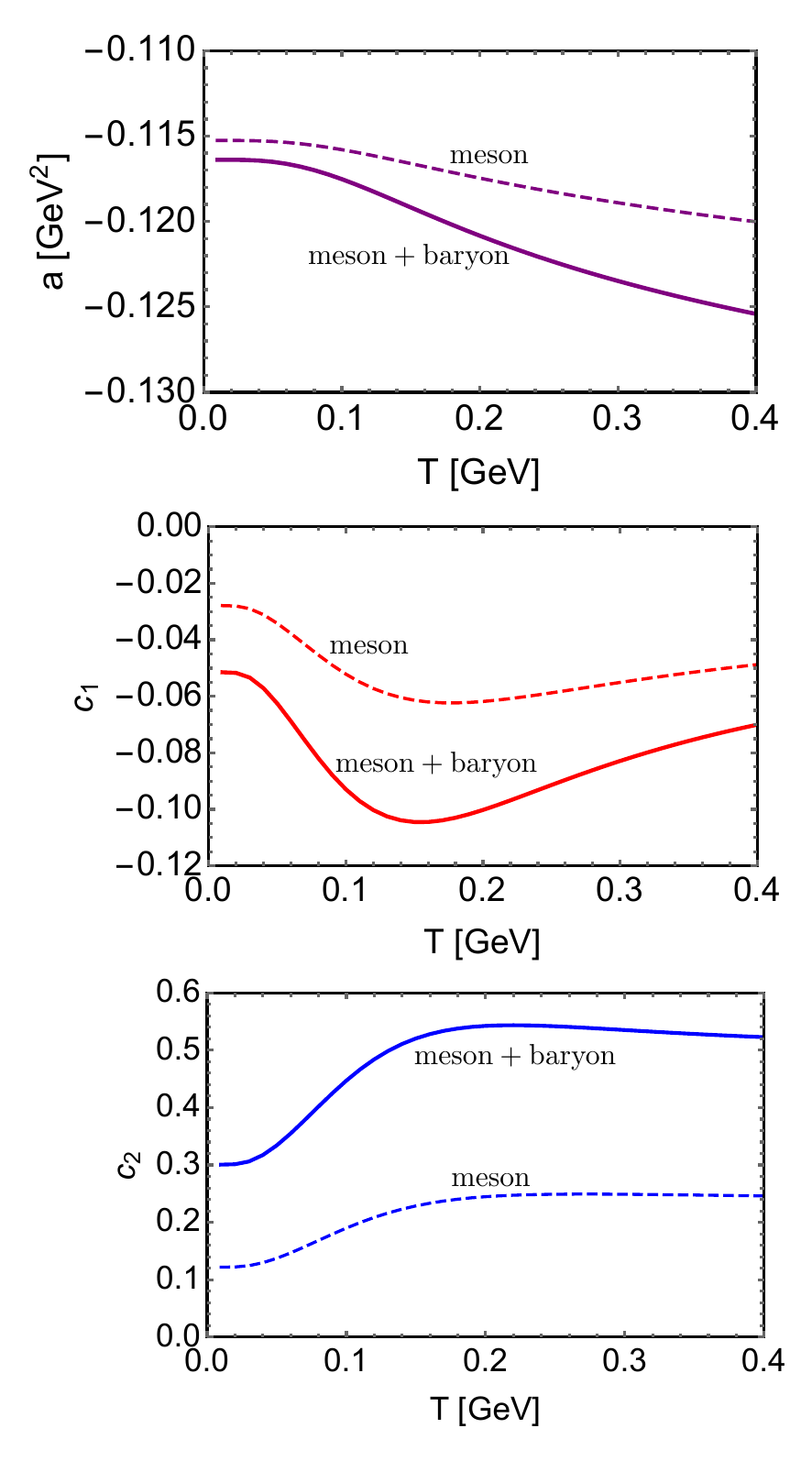}
\caption{$T$ dependencies of the anomaly coefficients $a$, $c_1$ and $c_2$ evaluated by the full (solid) and truncated (dashed) analyses, with $\mathring{M}_\eta=0.5$ GeV. }
\label{fig:BaryonContribution}
\end{figure}

\begin{figure}[t]
\centering
\hspace*{-0.5cm} 
\includegraphics*[scale=0.5]{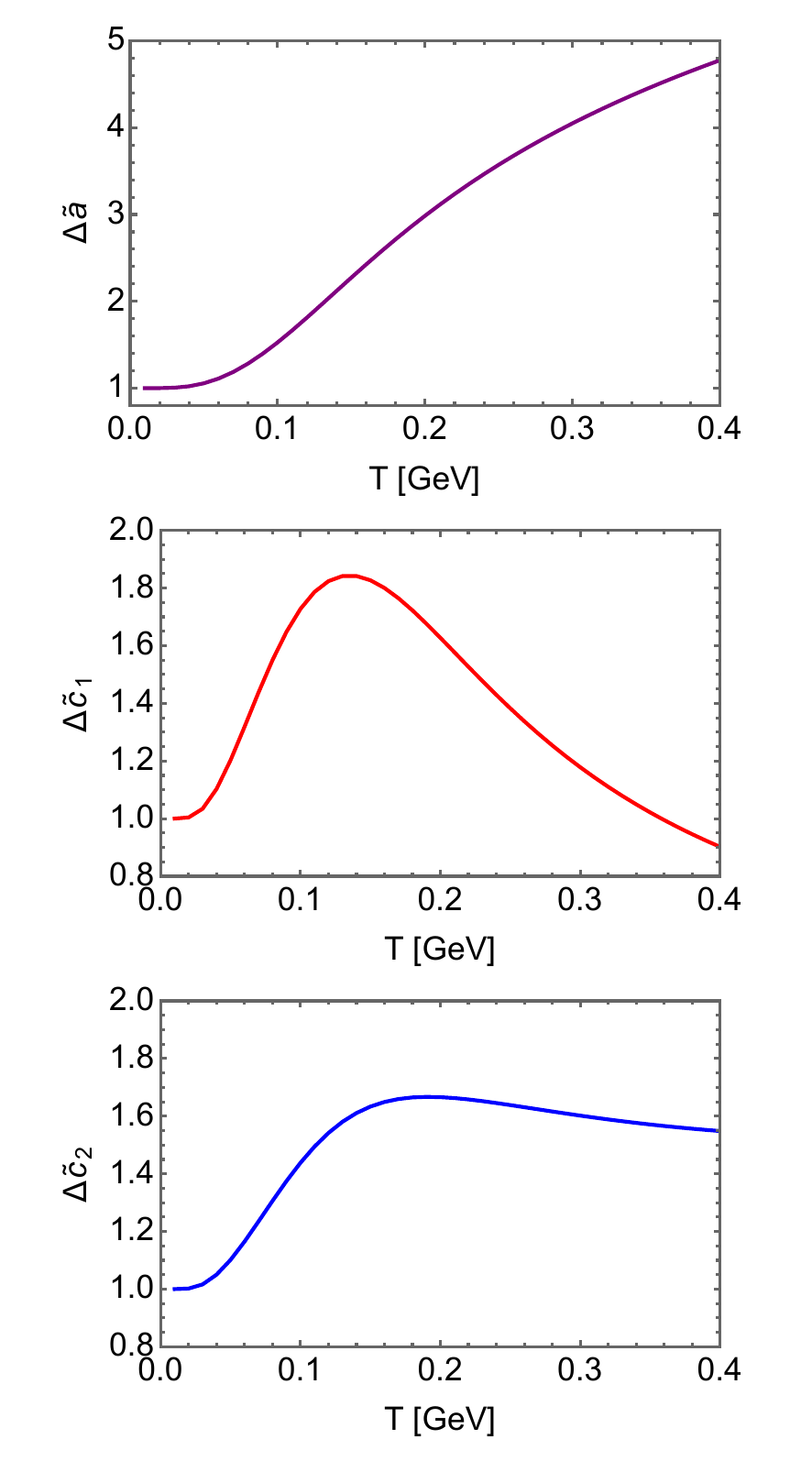}
\caption{$T$ dependencies of $\Delta\tilde{a}$, $\Delta\tilde{c}_1$, and $\Delta\tilde{c}_2$ defined in Eq.~(\ref{DeltaDef}), with $\mathring{M}_\eta=0.5$ GeV. }
\label{fig:BaryonContribution2}
\end{figure}

\section{Importance of baryonic contributions}
\label{sec:Baryons}

In the previous section, modifications of the hadron masses and anomaly coefficients were examined. In this section, we shed light on the role of baryonic contributions in the FRG framework to gain insight into fluctuations specific to QC$_2$D, where diquark baryons are also counted as hadrons. Here, we concentrate on the case where $\mathring{M}_\eta=0.5$ GeV.

In order to study baryonic contributions, on top of the full solution, we also solve the flow equations containing solely mesonic contributions (e.g., $\partial_k m^2_k = \tilde{\partial}_k\hat{V}^{I_1}_{M,k}$ for $m^2_k$) with inputs of $\mathring{M}_\pi=0.14$ GeV, $\mathring{M}_\eta=0.5$ GeV, $\lambda_{1,\Lambda}=\lambda_{2,\Lambda}/4$ and $c_{1,\Lambda}=c_{2,\Lambda}=0$. The value of $\lambda_{1,\Lambda}$ is adjusted such that $\mathring{M}_\sigma\approx 0.18$ GeV.
Then, we compare the cases with and without baryons.

The parameters $m^2_k$, $a_k$ and $\lambda_{1,k}$ at $k=0$ are roughly determined by the masses of the pion, $\eta$, and $\sigma$ mesons, respectively, and, hence those values in the truncated and in the full analysis do not differ much. On the other hand, $c_{1,k}$ and $c_{k,2}$ are simply fixed to vanish at $k=\Lambda_{\rm UV}$, and, thus, their scale evolutions are expected to be a useful measure to see the differences between the two approaches. Depicted in Fig.~\ref{fig:AnomalyFlowK} are the resultant $k$ dependencies for $c_{1,k}$ and $c_{2,k}$ at $T=0$, evaluated separately for the full solution and that containing only mesonic contributions. This figure clearly shows that the evolutions of $c_{1,k}$ and $c_{2,k}$ start to deviate in the two cases, soon after departing from $k=\Lambda_{\rm UV}$. As a result, at $k=0$, the coefficients containing both mesonic and baryonic contributions are significantly magnified compared to the truncated case, indicating the important role of baryonic fluctuations. Such enhancements can be simply understood by an increase in the degrees of freedom in the flows.

Finally, we also depict in Figs.~\ref{fig:BaryonContribution} and~\ref{fig:BaryonContribution2} the temperature dependencies of the anomaly coefficients at $k=0$, evaluated using the full and the truncated analyses. In Fig.~\ref{fig:BaryonContribution2}, for instance, $\Delta\tilde{a}$ is defined by
\begin{eqnarray}
\Delta\tilde{a} &\equiv&  \frac{\Delta a}{\Delta a|_{T=0}} \ , \label{DeltaDef}
\end{eqnarray}
where
\begin{eqnarray}
\Delta a = a|_{\rm meson+baryon}-a|_{\rm meson},
\end{eqnarray}
measuring differences with and without baryonic contributions. The same applies to $\Delta\tilde{c}_1$ and $\Delta\tilde{c}_2$. These figures also show the importance of baryonic contributions at any temperature with regards to the anomaly coefficients.

\section{Conclusions}
\label{sec:Conclusions}
In this paper, we have investigated modifications of $U(1)_A$ anomaly effects at finite temperature in QC$_2$D with two flavors, by means of the FRG technique. We have adopted the LSM developed in Ref.~\cite{Suenaga:2022uqn} to properly take into account fluctuations of the low-lying scalar and pseudoscalar mesons, as well as baryons, based on a linear representation of the Pauli-G\"{u}rsey $SU(4)$ symmetry.

As a result, we have found that quartic couplings of the LSM need to be suppressed, in order to properly reproduce a reasonable reduction of the chiral condensate at finite temperature. Accordingly, mass differences between the chiral partners $\sigma$ and $\pi$, or $a_0$ and $\eta$, are always found to be tiny, regardless of the $U(1)$ axial anomaly strength, i.e., the mass of the $\eta$ meson. At higher temperatures, mass degeneracies of the chiral partners in accordance with chiral symmetry restoration are realized. We note that masses of the $0^+$ and $0^-$ (anti)diquark baryons degenerate with those of $\pi$ and $a_0$, respectively, owing to the Pauli-G\"{u}rsey $SU(4)$ symmetry. The topological susceptibility at high temperature is found to be suppressed due to the reduction of chiral condensate.

Once we enter the finite-temperature regime, all of the couplings responsible for the $U(1)$ axial anomaly become magnified, and, hence we have concluded that due to hadronic fluctuations, the axial anomaly is enhanced in a hot medium for QC$_2$D. Similar anomaly enhancements were also discussed in three-color QCD analyses~\cite{Fejos:2016hbp}. At high temperatures, the increase in the quartic couplings is gradually suppressed and eventually diminishes. Our results also confirm that baryonic fluctuations play a substantial role in the enhancements.

The thermal properties of QC$_2$D can be tested via lattice QCD simulations; therefore, we expect that the present analysis will provide useful insights into hadron masses, particularly from the perspective of chiral symmetry restoration and $U(1)_A$ anomaly effects in a hot medium, for future simulations with a light pion mass.

We note that, while the $U(1)_A$ anomaly effects are enhanced at finite temperature within the hadronic model, LSM, it is known that at sufficiently high temperature and/or density, instantons melt due to the Debye screening of electric gluons, for which the $U(1)_A$ symmetry is {\it effectively} restored~\cite{Gross:1980br,Rapp:1999qa}. In order to smoothly connect those two behaviors, it would be necessary to implement a crossover transition to quark matter, but it is beyond the scope of the present study.

We have seen that, unlike the anomalous couplings, the topological susceptibility decreases monotonically with increasing temperature. This behavior is analogous to the three-flavor scenario and suggests that $\chi_{\rm top}$ may not be the most reliable indicator of the strength of the anomaly at finite temperature. The primary reason for the monotonic decrease of $\chi_{\rm top}$ is its close connection with the chiral condensate [cf. Eq.~(\ref{ChiTopRed})]. As the condensate melts, $\chi_{\rm top}$ also decreases, even though all anomalous interactions between the corresponding hadrons increase with temperature. This indicates that, when performing lattice simulations of the system, one must carefully choose observables that appropriately capture the anomalous breaking of $U(1)_A$.

Our analysis can also be applied to the case of a finite chemical potential. In the low-temperature and high-density regime, the emergence of a baryon superfluid phase, characterized by the condensation of positive-parity diquarks, has been demonstrated in lattice QCD simulations~\cite{Boz:2019enj,Buividovich:2020dks,Astrakhantsev:2020tdl,Iida:2024irv,Braguta:2023yhd} and model analyses~\cite{Kogut:2000ek,Ratti:2004ra,Sun:2007fc,Suenaga:2022uqn,Suenaga:2025sln}. In this phase, in particular, the possibility of enhanced $U(1)_A$ anomaly effects has been discussed by comparing the lattice data with LSM analyses of the hadron mass spectrum and topological susceptibilities ~\cite{Suenaga:2022uqn,Kawaguchi:2023olk}. Hence, investigating the fate of the $U(1)_A$ anomaly effects in the superfluid phase is considered an important direction for future work.

Finally, we provide some comments regarding diquarks. In three-color QCD, the chiral dynamics of diquarks is reflected in the spectrum of singly heavy baryons, which consist of a heavy quark $Q$ and a light diquark, either [$qq$] or [$sq$] ($q=u,d$). In this context, it is suggested that the $U(1)_A$ anomaly effects lead to the so-called {\it inverse mass hierarchy} of $M_{\Xi_Q(1/2)^-}<M_{\Lambda_Q(1/2^-)}$, or in terms of diquarks, $M_{[sq](0^-)}<M_{[qq](0^-)}$, despite their quark contents~\cite{Harada:2019udr,Suenaga:2023tcy,Suenaga:2024vwr}. QC$_2$D also serves as a useful testing ground to explore possible diquark properties. The present LSM analysis can be straightforwardly extended to the $N_f=2+1$ case, where both [$qq$] and [$sq$] diquarks are included. Therefore, it is worthwhile to investigate the fate of the inverse mass hierarchy at finite temperature, particularly in relation to possible anomaly enhancements. This is left for future work.

\section*{Acknowledgments}

G.F. was supported by the Hungarian National Research, Development, and Innovation Fund under Project No. FK142594. D.S. was supported by Grants-in-Aid for Scientific Research No. 23K03377, No. 23H05439 and No. 25K17386, from Japan Society for the Promotion of Science. G.F. expresses his gratitude for the warm hospitality of the Kobayashi-Maskawa Institute for the Origin of Particles and the Universe at Nagoya University during his stay. D.S. thanks the Young Researchers Overseas Program in Science of Nagoya University for supporting his stay at E\"{o}tv\"{o}s Lor\'{a}nd University. D.S. also thanks E\"{o}tv\"{o}s Lor\'{a}nd University for providing a comfortable research environment.

\appendix

\section{Derivation of flow equations for $\lambda_{1k}$, $\lambda_{2k}$, $c_{1k}$, and $c_{2k}$}
\label{sec:kTildeDerivative}

In Sec.~\ref{sec:EffectiveAction}, it is demonstrated that the flow equations for $m_k$ and $a_k$ can be obtained via evaluating part of the effective potential that is quadratic in the field variables. In this appendix, we show derivations of flow equations for the remaining quartic couplings: $\lambda_{1k}$, $\lambda_{2k}$, $c_{1k}$, and $c_{2k}$.

According to the expansion~(\ref{OmegaExpanded}), the ${\cal O}(\Sigma^4)$ part of the potential flow reads
\begin{eqnarray}
\partial_k V_k^{(4)} \equiv - \frac{T}{4}\sum_n\int\frac{d^3p}{(2\pi)^3}\tilde{\partial}_k{\rm tr}\Big[{\cal D}_{0,k}V_{(2)k}''{\cal D}_{0,k}V_{(2)k}''\Big]\ , \nonumber\\\label{Vk4Def}
\end{eqnarray}
where the definitions of ${\cal D}_{0,k}$ and $V_{(2)k}''$ are given by Eq.~(\ref{D0kInverse}); see also the explanations below Eq.~(\ref{OmegaExpanded}). As long as we stick to the zero chemical potential, ${\cal D}_{0,k}$ is diagonal and no Matsubara frequencies $\omega_n$ appear in any numerator; thus, the evaluation of the rhs of Eq.~(\ref{Vk4Def}) can be performed in a straightforward way. Making use of the identity
\begin{eqnarray}
\frac{1}{(i\omega_n)^2-\epsilon^2} = \frac{1}{2\epsilon}\left(\frac{1}{i\omega_n-\epsilon}-\frac{1}{i\omega_n+\epsilon}\right)
\end{eqnarray}
to decompose the propagators, the Matsubara summations are readily done by virtue of ($\omega_n=2\pi nT$)
\begin{eqnarray}
&& T\sum_n\frac{1}{(i\omega_n-\epsilon_1)(i\omega_n-\epsilon_2)} = \frac{f_B(\epsilon_2)-f_B(\epsilon_1)}{\epsilon_1-\epsilon_2}\ , \nonumber\\
&&  T\sum_n\frac{1}{(i\omega_n-\epsilon)^2} = \lim_{\epsilon_2\to \epsilon}\frac{f_B(\epsilon_2)-f_B(\epsilon)}{\epsilon-\epsilon_2} = -\frac{df_B(\epsilon)}{d\epsilon}\ , \nonumber\\
\end{eqnarray}
with the Bose-Einstein distribution function $f_B(\epsilon) = 1/({\rm e}^{\epsilon/T}-1)$. Then, the flow of $V_k^{(4)}$ is evaluated as
\begin{widetext}
\begin{eqnarray}
\partial_k V_k^{(4)}= \left( \partial_k\hat{V}_{M,k}^{I_1^2} + \partial_k\hat{V}_{B,k}^{I_1^2} \right)I_1^2 + \left(  \partial_k\hat{V}_{M,k}^{I_2} + \partial_k\hat{V}_{B,k}^{I_2}  \right)I_2 + \left(  \partial_k\hat{V}_{M,k}^{I_A^2} + \partial_k\hat{V}_{B,k}^{I_A^2}  \right)I_A^2 + \left(  \partial_k\hat{V}_{M,k}^{I_1I_A} + \partial_k\hat{V}_{B,k}^{I_1I_A}  \right)I_1I_A\ , \nonumber\\
\end{eqnarray}
where
\begin{eqnarray}
\partial_k\hat{V}_{M,k}^{I_1^2} &=&-\frac{1}{4} \int\frac{d^3p}{(2\pi)^3}\tilde{\partial}_k \nonumber\\
&&\Bigg(\frac{16 (c_{1,k}^2 + 3 c_{1,k} c_{2,k} + 3 c_{2,k}^2 + 4 c_{1,k} \lambda_{1,k} + 9 c_{2,k} \lambda_{1,k} + 7 \lambda_{1,k}^2) + 12 (c_{2,k} + 2 \lambda_{1,k}) \lambda_{2,k}+ 3 \lambda_{2,k}^2}{8(E_{k}^+)^{2}} F_1(E_k^+) \nonumber\\
 && +\frac{16 (c_{1,k}^2 - 3 c_{1,k} c_{2,k} + 3 c_{2,k}^2 + 4 c_{1,k} \lambda_{1,k} - 9 c_{2,k} \lambda_{1,k} + 7 \lambda_{1,k}^2) - 12 (c_{2,k} - 2 \lambda_{1,k}) \lambda_{2,k} + 3 \lambda_{2,k}^2}{8(E_{k}^-)^2} F_1(E_k^-) \nonumber\\
 && + \frac{16 c_{1,k}^2 - 32 c_{1,k} \lambda_{1,k} + 16 \lambda_{1,k}^2 + 3  \lambda_{2,k}^2}{4E_{k}^+E_{k}^-}F_2(E_k^+,E_k^-) \Bigg)\ ,  \label{VI12Mk}\\
\partial_k\hat{V}_{B,k}^{I_1^2} &=&-\frac{1}{4} \int\frac{d^3p}{(2\pi)^3}\tilde{\partial}_k\Bigg(\frac{(2 c_{2,k} + 4 \lambda_{1,k} + \lambda_{2,k})^2 }{4(E_{k}^+)^2}F_1(E_k^+) +\frac{(2 c_{2,k} - 4 \lambda_{1,k} - \lambda_{2,k})^2}{4(E_{k}^-)^2}F_1(E_k^-) + \frac{\lambda_{2,k}^2}{2E_{k}^+E_{k}^-}F_2(E_k^+,E_k^-) \Bigg)\ , \nonumber\\ \\
\partial_k\hat{V}_{M,k}^{I_2} &=& -\frac{1}{4} \int\frac{d^3p}{(2\pi)^3}\tilde{\partial}_k\Bigg(\frac{4(c_{1,k}+c_{2,k}+\lambda_{1,k})\lambda_{2,k}}{(E_{k}^+)^2} F_1(E_k^+) + \frac{4(c_{1,k}-c_{2,k}+\lambda_{1,k})\lambda_{2,k}}{(E_{k}^-)^2} F_1(E_k^-) \nonumber\\
&& \hspace{2.7cm}  - \frac{2(8c_{1,k}-8\lambda_{1,k}-\lambda_{2,k})\lambda_{2,k}}{E_{k}^+E_{k}^-}F_2(E_k^+,E_k^-) \Bigg)\ ,\\ 
\partial_k\hat{V}_{B,k}^{I_2} &=& -\frac{1}{4}\int\frac{d^3p}{(2\pi)^3}\tilde{\partial}_k \left(-\frac{2\lambda_{2,k}^2}{E_{k}^+E_{k}^-}F_2(E_k^+,E_k^-)\right)\ , \\
\partial_k\hat{V}_{M,k}^{I_A^2} &=& -\frac{1}{4} \int\frac{d^3p}{(2\pi)^3}\tilde{\partial}_k\nonumber\\
&&\Bigg(\frac{16 (7 c_{1,k}^2 + 9 c_{1,k} c_{2,k} + 3 c_{2,k}^2 + 4 c_{1,k} \lambda_{1,k} + 3 c_{2,k} \lambda_{1,k} + \lambda_{1,k}^2) - 12 (2 c_{1,k} + c_{2,k}) \lambda_{2,k} + 3 \lambda_{2,k}^2}{8(E^+_{k})^2} F_1(E_k^+) \nonumber\\
&& +\frac{16 (7 c_{1,k}^2 - 9 c_{1,k} c_{2,k} + 3 c_{2,k}^2 + 4 c_{1,k} \lambda_{1,k} - 3 c_{2,k} \lambda_{1,k} + \lambda_{1,k}^2) + 12 (-2 c_{1,k} + c_{2,k}) \lambda_{2,k} + 3 \lambda_{2,k}^2}{8(E_{k}^-)^2}F_1(E_k^-)\nonumber\\
 && - \frac{16c_{1,k}^2-32c_{1,k}\lambda_{1,k}+16\lambda_{1,k}^2+3\lambda_{2,k}^2}{4E_{k}^+E_{k}^-}F_2(E_k^+,E_k^-)  \Bigg)\ , 
 \end{eqnarray}
\begin{eqnarray}
\partial_k\hat{V}_{B,k}^{I_A^2}  &=& -\frac{1}{4} \int\frac{d^3p}{(2\pi)^3}\tilde{\partial}_k\nonumber\\
&&\Bigg(\frac{(4 c_{1,k} + 2 c_{2,k} - \lambda_{2,k})^2 }{4(E_{k}^+)^2}F_1(E_k^+)  +\frac{(4 c_{1,k} - 2 c_{2,k} - \lambda_{2,k})^2}{4(E_{k}^-)^2}F_1(E_k^-) - \frac{\lambda_{2,k}^2}{2E_{k}^+E_{k}^-}F_2(E_k^+,E_k^-)  \Bigg)\ , \\
\partial_k\hat{V}_{M,k}^{I_1I_A} &=&-\frac{1}{4} \int\frac{d^3p}{(2\pi)^3}\tilde{\partial}_k\nonumber\\
&&\Bigg(\frac{16 (2 c_{1,k}^2 + 6 c_{1,k} c_{2,k} + 3 c_{2,k}^2 + 8 c_{1,k} \lambda_{1,k} + 6 c_{2,k} \lambda_{1,k} +  2 \lambda_{1,k}^2) + 12 (c_{1,k} - \lambda_{1,k}) \lambda_{2,k} - 3 \lambda_{2,k}^2}{4(E_{k}^+)^2} F_1(E_k^+)  \nonumber\\
 && -\frac{16 (2 c_{1,k}^2 - 6 c_{1,k} c_{2,k} + 3 c_{2,k}^2 + 8 c_{1,k} \lambda_{1,k} - 6 c_{2,k} \lambda_{1,k} +  2 \lambda_{1,k}^2) + 12 (c_{1,k} - \lambda_{1,k}) \lambda_{2,k} - 3 \lambda_{2,k}^2}{4(E_{k}^-)^2} F_1(E_k^-)\Bigg)\ , \nonumber\\
\end{eqnarray}
and
\begin{eqnarray}
&&\hspace{-4.5cm}\partial_k\hat{V}_{B,k}^{I_1I_A} = -\frac{1}{4} \int\frac{d^3p}{(2\pi)^3}\tilde{\partial}_k\Bigg(\frac{(4 c_{1,k} + 2 c_{2,k} - \lambda_{2,k}) (2 c_{2,k} + 4 \lambda_{1,k} + \lambda_{2,k}) }{2(E_{k}^+)^2}F_1(E_k^+)  \nonumber\\
&& + \frac{(4 c_{1,k} - 2 c_{2,k} - \lambda_{2,k}) (2 c_{2,k} - 4 \lambda_{1,k} - \lambda_{2,k}) }{2(E_{k}^-)^2}F_1(E_k^-)\Bigg)  \label{VI1IABk}
\end{eqnarray}
\end{widetext}
are representing contributions of mesonic and baryonic fluctuations to the respective coefficients. In addition, we note that $E_k^\pm = \sqrt{{\bm p}^2+R_k({\bm p})+(m^\pm_{{\rm eff},k})^2}$ with $(m^\pm_{{\rm eff},k})^2 = m_k^2\pm a_k $, and $\tilde{\partial}_k$ is a differential operator acting only on the regulator $R_k$. In the above equations, we have also defined
\begin{eqnarray}
F_1(\epsilon) &\equiv& \frac{1}{2\epsilon}\Big(1+2f_B(\epsilon)\Big)-\frac{d f_B(\epsilon)}{d\epsilon} \ , \nonumber\\
F_2(\epsilon_1,\epsilon_2) &\equiv& \frac{1}{\epsilon_1+\epsilon_2}\Big(1+f_B(\epsilon_1)+f_B(\epsilon_2)\Big) \nonumber\\
&& -\frac{f_B(\epsilon_1) - f_B(\epsilon_2)}{\epsilon_1-\epsilon_2}.
\end{eqnarray}
Here, the first terms in $F_1(\epsilon)$ and $F_2(\epsilon_1,\epsilon_2)$ stand for particle-antiparticle contributions, while the second ones stand for particle-hole contributions, respectively.

Using Eqs.~(\ref{VI12Mk})--(\ref{VI1IABk}), flow equations for the quartic couplings are obtained by
\begin{eqnarray}
\partial_k \lambda_{1k} &=& \partial_k\hat{V}^{I_1^2}_{M,k} + \partial_k\hat{V}^{I_1^2}_{B,k}\ , \nonumber\\
\partial_k \lambda_{2k} &=& \partial_k\hat{V}^{I_2}_{M,k} + \partial_k\hat{V}^{I_2}_{B,k}\ , \nonumber\\
\partial_k c_{1k} &=& \partial_k\hat{V}^{I_A^2}_{M,k} + \partial_k\hat{V}^{I_A^2}_{B,k}\ , \nonumber\\
\partial_k c_{2k} &=& \partial_k\hat{V}^{I_1I_A}_{M,k} + \partial_k\hat{V}^{I_1I_A}_{B,k}\ .
\end{eqnarray}

The $\tilde{\partial}_k$ derivatives are evaluated using
\begin{eqnarray}
&& \int\frac{d^3p}{(2\pi)^3}\tilde{\partial}_k{\cal F}_k(E_k^+,E_{k}^-) \nonumber\\
&=& \int\frac{d^3p}{(2\pi)^3}\theta(k^2-{\bm p}^2)\nonumber\\
&&\ \ \ \  \times \sum_{\alpha=\pm}\left[\frac{k}{E_{k}^\alpha}\frac{\partial{\cal F}_k(E_{k}^+, E_{k}^-)}{\partial E_{k}^\alpha}\right]_{E_{k}^\pm= \sqrt{k^2+(m^\pm_{{\rm eff},k})^2}}  \nonumber\\ \label{KDFormula}
&=& \frac{k^4}{6\pi^2}\sum_{\alpha=\pm}\left[\frac{1}{E_{k}^\alpha}\frac{\partial{\cal F}_k(E_{k}^+,E_{k}^-)}{\partial E_{k}^\alpha}\right]_{E_{k}^\pm= \sqrt{k^2+(m^\pm_{{\rm eff},k})^2}}\, , \nonumber\\\
\end{eqnarray}
where
\begin{eqnarray}
\frac{\partial E_{k}^\pm}{\partial k} &=& \frac{\delta(k^2-{\bm p}^2)(k^2-{\bm p}^2)}{E_{k}^\pm} + \frac{\theta(k^2-{\bm p}^2)k}{E_{k}^\pm} \nonumber\\
&\overset{p\, {\rm integral}}{\to}& \frac{\theta(k^2-{\bm p}^2)k}{\sqrt{k^2+(m^\pm_{{\rm eff},k})^2} } 
\end{eqnarray}
is adopted.

\section{Hadron mass formulas}
\label{sec:HadronMassFormula}

Here, we show analytic formulas for the hadron masses, which are the eigenvalues of the second derivative matrix (\ref{VkDerivative}) of the potential~(\ref{VkAnsatz}) in the homogeneous background, $\sigma_0$. Using Eq.~(\ref{VkAssume}), the resultant formulas read
\begin{eqnarray}
M_{\eta,k}^2 &=& m_k^2-a_k+\lambda_{1,k}\sigma_0^2-c_{1,k} \sigma_0^2 \nonumber\\
M_{\pi,k}^2 &=& M_{B,k}^2= M_{\bar{B},k}^2 \nonumber\\
&=& m_k^2+a_k+\lambda_{1,k}\sigma_0^2 +(c_{1,k}+c_{2,k})\sigma_0^2 \ ,  \label{PiMass}
\end{eqnarray}
and
\begin{eqnarray}
M_{\sigma,k}^2 &=& m_k^2+a_k+3\lambda_{1,k}\sigma_0^2 + 3(c_{1,k}+c_{2,k})\sigma_0^2 \ ,\nonumber\\
M_{a_0,k}^2 &=& M_{B',k}^2 = M^2_{\bar{B}',k} \nonumber\\
&=& m_k^2-a_k+\left(\lambda_{1,k}+\frac{\lambda_{2,k}}{2}\right)\sigma_0^2 -c_{1,k}\sigma_0^2 \ . \nonumber\\ \label{SigmaMass}
\end{eqnarray}

The gap equation that determines $\sigma_0$ is
\begin{eqnarray}
 m_k^2+a_k+\lambda_{1,k}\sigma_0^2 +(c_{1,k}+c_{2,k})\sigma_0^2 = \frac{\sqrt{2}m_q\bar{c}}{\sigma_0} \ , \label{GapEq}
\end{eqnarray}
where $m_q\bar{c}$ does not depend on the separation scale $k$. Thus, one can find
\begin{eqnarray}
M^2_{\pi,k} \sigma_0 = \sqrt{2}m_q\bar{c} = ({\rm const})\ , \label{PiMassRelation}
\end{eqnarray}
at any $k$ and temperature. In the chiral limit $m_q\bar{c}\to0$, inserting Eq.~(\ref{GapEq}) into Eq.~(\ref{PiMass}) leads to $M_{\pi,k}^2=M_{B(\bar{B}),k}^2=0$, which yields the appearance of five Nambu-Goldstone bosons corresponding to the breaking pattern of $SU(4)\to Sp(4)$.

From Eqs.~(\ref{PiMass}) and~(\ref{SigmaMass}), one can confirm
\begin{eqnarray}
M_{\sigma,k}^2- M_{\pi,k}^2 &=& 2(\lambda_{1,k}-c_{1,k}-c_{2,k})\sigma_0^2 \ ,\nonumber\\
M_{a_0,k}^2- M_{\eta,k}^2 &=& \frac{\lambda_{2,k}}{2}\sigma_0^2\ . \label{MassDiff}
\end{eqnarray}
Hence, when chiral symmetry is restored, $\sigma_0\to0$, the chiral partner structures $M_{\sigma,k}=M_{\pi,k}$ and $M_{a_0,k}=M_{\eta,k}$ are realized, as they should.

Under the assumption that $c_{1,k}$ and $c_{2,k}$ are negligible, the mass difference between $\eta$ and $\pi$ is approximated as
\begin{eqnarray}
M^2_{\eta,k} &=& -2a_k-(2c_{1,k}+c_{2,k})\sigma_0^2 \nonumber\\
&\approx& M^2_{\pi,k}  -2a_k\ , \label{PiEtaMass}
\end{eqnarray}
indicating that the $\eta$ mass is dominantly generated by the quadratic anomalous term.

\bibliography{reference}

\end{document}